\documentclass[twocolumn]{aastex631}

\shorttitle{SN~2023wrk}
\shortauthors{Liu J. et al.}
\graphicspath{{./}{}}
\begin{document}

\title{Early-Time Observations of SN~2023wrk: A Luminous Type Ia Supernova \\ with Significant Unburned Carbon in the Outer Ejecta}

\author[0009-0000-0314-6273]{Jialian Liu}
\affiliation{Physics Department, Tsinghua University, Beijing 100084, China}
\email{liu-jl22@mails.tsinghua.edu.cn}

\author[0000-0002-7334-2357]{Xiaofeng Wang}
\affiliation{Physics Department, Tsinghua University, Beijing 100084, China}

\email{wang\_xf@mail.tsinghua.edu.cn}

\author[0009-0004-9687-3275]{Cristina Andrade} % andra104@umn.edu
\affiliation{School of Physics and Astronomy, University of Minnesota, Minneapolis, MN 55455, USA\\}

\author[0000-0002-3906-0997]{Pierre-Alexandre Duverne} % duverne@apc.in2p3.fr
\affiliation{ Universit\'e Paris Cit\'e, CNRS, Astroparticule et Cosmologie, F-75013 Paris, France\\ }

\author{Jujia Zhang} % jujia@ynao.ac.cn
\affiliation{Yunnan Observatories, Chinese Academy of Sciences, Kunming 650216, China}
\affiliation{International Centre of Supernovae, Yunnan Key Laboratory, Kunming 650216, China}
\affiliation{Key Laboratory for the Structure and Evolution of Celestial Objects, Chinese Academy of Sciences, Kunming 650216, China}

\author{Liping Li} % liliping@ynao.ac.cn
\affiliation{Yunnan Observatories, Chinese Academy of Sciences, Kunming 650216, China}
\affiliation{International Centre of Supernovae, Yunnan Key Laboratory, Kunming 650216, China}
\affiliation{Key Laboratory for the Structure and Evolution of Celestial Objects, Chinese Academy of Sciences, Kunming 650216, China}

\author{Zhenyu Wang} % wangzhenyu@ynao.ac.cn
\affiliation{Yunnan Observatories, Chinese Academy of Sciences, Kunming 650216, China}
\affiliation{International Centre of Supernovae, Yunnan Key Laboratory, Kunming 650216, China}
\affiliation{Key Laboratory for the Structure and Evolution of Celestial Objects, Chinese Academy of Sciences, Kunming 650216, China}
\affiliation{University of Chinese Academy of Sciences, Beijing 100049, China }

\author[0000-0002-0284-0578]{Felipe Navarete} % felipe.navarete@noirlab.edu
\affiliation{SOAR Telescope/NSF's NOIRLab Avda Juan Cisternas 1500, 1700000, La Serena, Chile\\}

\author[0000-0003-4254-2724]{Andrea Reguitti} % andreareguitti@gmail.com
\affiliation{INAF -- Osservatorio Astronomico di Brera, Via E. Bianchi 46, 23807 Merate (LC), Italy}
\affiliation{INAF -- Osservatorio Astronomico di Padova, Vicolo dell'Osservatorio 5, 35122 Padova, Italy}

\author[0000-0003-2497-6334]{Stefan Schuldt} % 
\affiliation{Dipartimento di Fisica, Università degli Studi di Milano, via Celoria 16, I-20133 Milano, Italy}
\affiliation{INAF - IASF Milano, via A. Corti 12, I-20133 Milano, Italy}

\author[0000-0002-7714-493X]{Yongzhi Cai} % yzcai789@163.com
\affiliation{Yunnan Observatories, Chinese Academy of Sciences, Kunming 650216, China}
\affiliation{International Centre of Supernovae, Yunnan Key Laboratory, Kunming 650216, China}
\affiliation{Key Laboratory for the Structure and Evolution of Celestial Objects, Chinese Academy of Sciences, Kunming 650216, China}

\author{Alexei V. Filippenko} % alex@astro.berkeley.edu
\affiliation{Department of Astronomy, University of California, Berkeley, CA 94720-3411, USA}

\author{Yi Yang} % yiyangtamu@gmail.com
\affiliation{Physics Department, Tsinghua University, Beijing 100084, China}
\affiliation{Department of Astronomy, University of California, Berkeley, CA 94720-3411, USA}
%\affiliation{Bengier-Winslow-Robertson Postdoctoral Fellow}

\author{Thomas G. Brink} %tgbrink@berkeley.edu
\affiliation{Department of Astronomy, University of California, Berkeley, CA 94720-3411, USA}
%\affiliation{Draper-Wood-Robertson Specialist in Astronomy}

\author{WeiKang Zheng} %zwk@astro.berkeley.edu
\affiliation{Department of Astronomy, University of California, Berkeley, CA 94720-3411, USA}
%\affiliation{Bengier-Winslow-Eustace Specialist in Astronomy}

\author{Ali Esamdin} % aliyi@xao.ac.cn
%\email{aliyi@xao.ac.cn}
\affiliation{Xinjiang Astronomical Observatory, Chinese Academy of Sciences, Urumqi, Xinjiang, 830011, China}

\author{Abdusamatjan Iskandar} % abudu@xao.ac.cn
\affiliation{Xinjiang Astronomical Observatory, Chinese Academy of Sciences, Urumqi, Xinjiang, 830011, China}
\affiliation{School of Astronomy and Space Science, University of Chinese Academy of Sciences, Beijing 100049, China}

\author{Chunhai Bai}  % baichunhai@xao.ac.cn
\affiliation{Xinjiang Astronomical Observatory, Chinese Academy of Sciences, Urumqi, Xinjiang, 830011, China}

\author{Jinzhong Liu}  % liujinzh@xao.ac.cn
\affiliation{Xinjiang Astronomical Observatory, Chinese Academy of Sciences, Urumqi, Xinjiang, 830011, China}

\author[0000-0001-5879-8762]{Xin Li} % iodblll@126.com
\affiliation{Beijing Planetarium, Beijing Academy of Science and Technology, Beijing 100044, China}

\author{Maokai Hu} % kaihukaihu123@mail.tsinghua.edu.cn
\affiliation{Physics Department, Tsinghua University, Beijing 100084, China}

\author{Gaici Li} % lgc21@mails.tsinghua.edu.cn
\affiliation{Physics Department, Tsinghua University, Beijing 100084, China}

\author{Wenxiong Li} % li-wx15@tsinghua.org.cn
\affiliation{Natioal Astronomical Observatories of China, Chinese Academy of Sciences, Beijing, 100012, China}

\author{Xiaoran Ma} % maxr20@mails.tsinghua.edu.cn
\affiliation{Physics Department, Tsinghua University, Beijing 100084, China}

\author{Shengyu Yan} % yansy19@mails.tsinghua.edu.cn
\affiliation{Physics Department, Tsinghua University, Beijing 100084, China}

\author{Jun Mo} % moj20@mails.tsinghua.edu.cn
\affiliation{Physics Department, Tsinghua University, Beijing 100084, China}

\author[0000-0002-8904-3925]{Christophe Adami} % christophe.adami@lam.fr
\affiliation{Aix Marseille Univ, CNRS, CNES, LAM, Marseille, France\\}

\author[0009-0006-4358-9929]{Dalya Akl} % g00088720@aus.edu
\affiliation{Physics Department, American University of Sharjah, Sharjah, UAE\\ }

\author{Sarah Antier} % sarah.antier@oca.eu
\affiliation{Université Côte d’Azur, Observatoire de la Côte d’Azur, CNRS, Laboratoire J.-L. Lagrange, Boulevard de l’Observatoire, 06304 Nice, France\\}

\author[0000-0003-1523-4478]{Eric Broens} % eric.broens@skynet.be
\affiliation{KNC, Vereniging Voor Sterrenkunde, Balen-Neetlaan 18A, B-2400, Mol, Belgium\\ }

\author{Jean-Grégoire Ducoin} % ducoin@iap.fr
\affiliation{CPPM, Aix Marseille Univ, CNRS/IN2P3, CPPM, Marseille, France\\}

\author[0000-0002-9751-8089]{Eslam Elhosseiny} % eslam_elhosseiny@nriag.sci.eg
\affiliation{ National Research Institute of Astronomy and Geophysics (NRIAG), 1 El-marsad St., 11421 Helwan, Cairo, Egypt\\}

\author[0000-0002-0792-3719]{Thomas M. Esposito} % tesposito@seti.org
\affiliation{Department of Astronomy, University of California, Berkeley, CA 94720-3411, USA}

\author[0009-0005-4287-7198]{Michael Freeberg} % mdfreeb@gmail.com
\affiliation{KNC, AAVSO, Hidden Valley Observatory (HVO), Colfax, WI  54730, USA}
\affiliation{iTelescope, Utah Desert Remote Observatory, Beryl Junction, UT 84714, USA\\} 

\author{Priya Gokulass} % priyadass.94@gmail.com
\affiliation{Department of Physical Sciences, College of Arts and Sciences, Embry-Riddle Aeronautical University, Daytona Beach, FL 32114, USA\\}

\author{Patrice Hello} % hello@lal.in2p3.fr
\affiliation{IJCLab, Univ Paris-Saclay, CNRS/IN2P3, Orsay, France\\}

\author[0000-0003-0035-651X]{Sergey Karpov} % karpov@fzu.cz
\affiliation{CEICO, Institute of Physics of the Czech Academy of Sciences, Na Slovance 1999/2, CZ-182 21, Praha, Czech Republic\\}

\author[0000-0003-2629-1945]{Isabel Márquez} % isabel.marquez@iaa.csic.es
\affiliation{Instituto de Astrofísica de Andalucía (IAA-CSIC), Glorieta de la Astronomía, 18008 Granada, Spain} %%TNG

\author[0000-0002-0967-0006]{Martin Mašek} % cassi@astronomie.cz
\affiliation{FZU -- Institute of Physics, Czech Academy of Sciences, Na Slovance 1999/2, CZ-182 21, Praha, Czech Republic\\}

\author{Oleksandra Pyshna} % pyshnaya.sasha@gmail.com
\affiliation{The Astronomical observatory of Taras Shevchenko National University of Kyiv, Ukraine\\}

\author{Yodgor Rajabov} % y.rajabov94@bk.ru
\affiliation{Ulugh Beg Astronomical Institute, Uzbekistan Academy of Sciences, Astronomy Str. 33, Tashkent 100052, Uzbekistan\\}

\author{Denis Saint-Gelais} % denis_st-gelais@hotmail.fr
\affiliation{KNC, Sociedad Astronomica Queretana, Mexico, RAPAS Observatoire de Paris, France, Le club des Astronome Amateur de Rosemère du Québec, Canada}

\author[0009-0003-5793-4293]{Marc Serrau} % marc.serrau2@free.fr
\affiliation{KNC, Société Astronomique de France, Observatoire de Dauban, F-04150 Banon, France\\} 

\author[0000-0003-4503-7272]{Oleksii Sokoliuk} % oleksii.sokoliuk@mao.kiev.ua
\affiliation{ Astronomical Observatory of Taras Shevchenko National University of Kyiv, Observatorna Str. 3, Kyiv, 04053, Ukraine\\ }
\affiliation{ Main Astronomical Observatory of National Academy of Sciences of Ukraine, 27 Acad. Zabolotnoho Str., Kyiv, 03143, Ukraine\\}

\author[0000-0003-1423-5516]{Ali Takey} % ali.takey@nriag.sci.eg
\affiliation{ National Research Institute of Astronomy and Geophysics (NRIAG), 1 El-marsad St., 11421 Helwan, Cairo, Egypt\\}

\author{Manasanun Tanasan} % cchann491@gmail.com
\affiliation{National Astronomical Research Institute of Thailand (Public Organization), 260, Moo 4, T. Donkaew, A. Mae Rim, Chiang Mai, 50180, Thailand\\}

\author[0000-0003-1835-1522]{Damien Turpin} % damien.turpin@cea.fr
\affiliation{Universit\'e Paris-Saclay, Universit\'e Paris Cit\'e, CEA, CNRS, AIM, 91191, Gif-sur-Yvette, France\\}




\begin{abstract}
We present extensive photometric and spectroscopic observations of the nearby Type Ia supernova (SN) 2023wrk at a distance of about 40 Mpc. The earliest detection of this SN can be traced back to a few hours after the explosion. Within the first few days the light curve shows a bump feature, while the $B - V$ color is blue and remains nearly constant. The overall spectral evolution is similar to that of an SN 1991T/SN 1999aa-like SN~Ia, while the C\,{\sc ii} $\lambda6580$ absorption line appears to be unusually strong in the first spectrum taken at $t \approx -$15.4 days after the maximum light. This carbon feature disappears quickly in subsequent evolution but it reappears at around the time of peak brightness. The complex evolution of the carbon line and the possible detection of Ni\,{\sc iii} absorption around 4700~\AA\ and 5300~\AA\ in the earliest spectra indicate macroscopic mixing of fuel and ash. The strong carbon lines is likely related to collision of SN ejecta with unbound carbon, consistent with the predictions of pulsational delayed-detonation or carbon-rich circumstellar-matter interaction models. 
Among those carbon-rich SNe~Ia with strong C\,{\sc ii} $\lambda6580$ absorption at very early times, the line-strength ratio of C\,{\sc ii} to Si\,{\sc ii} and the $B-V$ color evolution are found to exhibit large diversity, which may be attributed to different properties of unbound carbon and outward-mixing $^{56}$Ni.
\end{abstract}

%\keywords{Type Ia supernovae; distance scale;....}


\section{INTRODUCTION} \label{sec:intro}
Type Ia supernovae (SNe~Ia; see, e.g., \citealt{1997ARA&A..35..309F} for a review of SN classification) are widely believed to arise from thermonuclear explosions of carbon-oxygen (C/O) white dwarfs (WDs) \citep{1997Sci...276.1378N,2000ARA&A..38..191H}. Thanks to the high peak luminosity and width-luminosity relation of their light curves (WLR, also dubbed the ``Phillips relation"; \citealt{1993ApJ...413L.105P}; \citealt{1999AJ....118.1766P}), SNe~Ia have been used as standardizable candles in cosmology, playing a critical role in the discovery of the accelerating expansion of the Universe \citep{1998AJ....116.1009R,1999ApJ...517..565P} and measurement of the Hubble constant (e.g., \citealt{2022ApJ...934L...7R}, \citealt{2023JCAP...11..046M}). 

About three decades ago, SNe~Ia were classified into spectroscopically peculiar and normal (dubbed as ``Branch-normal") SNe~Ia \citep{1993AJ....106.2383B}. Later, more diversity has been noticed for SNe~Ia, even among the normal ones. For example, \citet{2005ApJ...623.1011B} found that SNe~Ia could be divided into low-velocity-gradient (LVG) and high-velocity-gradient (HVG) objects according to the velocity gradient measured from the Si\,{\sc ii} $\lambda6355$ absorption, and this velocity gradient is not correlated with $\Delta m_{15}(B)$ \citep{1993ApJ...413L.105P} for normal SNe~Ia. \citet{2006PASP..118..560B} assigned SNe~Ia into four groups based on the strengths of their Si\,{\sc ii} lines:  the ``shallow-silicon" group with a weak Si\,{\sc ii} $\lambda6355$ line, the ``cool" group with a high line-strength ratio of Si\,{\sc ii} $\lambda5972$ to Si\,{\sc ii} $\lambda6355$, the ``broad-line" group with strong Si\,{\sc ii} $\lambda6355$, and the ``core-normal" group with a moderate Si\,{\sc ii} line strength and a high degree of spectral homogeneity. Based on the Si\,{\sc ii} $\lambda6355$ velocity, \citet{2009ApJ...699L.139W} divided normal SNe~Ia into two categories: high velocity (HV) and normal velocity (NV), which may arise from different birth environments \citep{2013Sci...340..170W}. %The precision of SN~Ia distances may be improved after including the more-uniform subclass \citep{2009ApJ...699L.139W}.
\citet{2009ApJ...699L.139W} found an apparent dichotomy in the host 
galaxy reddening law ($R_V$ = 2.4 for "NV" SNe and $R_V$ = 1.6 for "HV" SNe), and concluded that 
applying such an absorption-correction dichotomy could significantly reduce the peak luminosity dispersion.
Some subtypes of SNe~Ia with peculiar properties compared with normal objects have also been found, such as the luminous SN~1991T-like or SN~1999aa-like (hereafter 91T/99aa-like, which belong to the Branch ``shallow-silicon" subclass) objects \citep{1992ApJ...384L..15F, 1992AJ....103.1632P,2000ApJ...539..658K}, the subluminous SN~1991bg-like (hereafter 91bg-like, which belong to the Branch ``cool" subclass) objects \citep{1992AJ....104.1543F,1993AJ....105..301L}, the super-Chandrasekhar-mass SN~2003fg-like objects \citep{2006Natur.443..308H}, and the moderately declining, subluminous SN~2002es-like objects \citep{2012ApJ...751..142G}. 

Among different subtypes of SNe~Ia, the 91T/99aa-like objects are characterized by prominent Fe\,{\sc ii/iii} absorption and weak Si\,{\sc ii} $\lambda6355$ absorption in the early-time spectra; 
91T-like SNe are systematically more luminous than 99aa-like and ``normal" SNe Ia with similar decline rates \citep{2021ApJ...912...71B, 2022ApJ...938...83Y, 2022ApJ...938...47P}.
This has been explained by more complete burning, more nickel synthesized in their explosions, and the resulting higher photospheric temperature \citep{2014MNRAS.445..711S}.
\citet{2022ApJ...938...47P} also emphasized that the post-peak decline rate alone cannot be used to differentiate between 91T-like, 99aa-like, and luminous SNe Ia with `normal" spectra.
Although the fraction of 91T/99aa-like SNe Ia in the local universe is estimated to be low ($\sim$4\% in the ASASSN survey \citep{2024MNRAS.530.5016D} and $\sim$8\% in X. Ma et al. (2024 in prep.)),
this fraction could increase in the distant universe owing to a shorter delay time of formation \citep{2023arXiv231103473C}. Moreover, the 91T/99aa-like objects could extend the cosmological sample to higher redshifts  \citep{2022ApJ...938...83Y} because of their relatively high luminosity. However, it is still debated whether 91T/99aa-like objects and normal subclasses arise from a common progenitor system and explosion mechanism and follow exactly the same WLR. The spectra of 91T/99aa-like SNe~Ia progressively resemble those of the normal ones around and beyond the time of peak brightness. Moreover, SN~1999aa exhibits some properties intermediate between those of SN~1991T and normal SNe~Ia, and SN~1999aa is more similar to normal SNe~Ia at earlier phases and shows prominent Ca\,{\sc ii} H\&K lines that are not seen in SN~1991T \citep{2004AJ....128..387G}. But these luminous SNe~Ia could introduce systematic biases when inferring cosmological distances with them \citep{2012ApJ...757...12S}.
 
Although having relatively high peak luminosities and slowly evolving light curves, there have not been many observations of 91T/99aa-like objects at very early times, which are important for constraining their progenitor systems and explosion mechanisms.  Exceptions to this include iPTF14bdn \citep{2015ApJ...813...30S} and
iPTF16abc \citep{2018ApJ...852..100M, 2024MNRAS.529.3838A}. But besides the weak Si\,{\sc ii} and strong Fe\,{\sc iii} absorption, iPTF16abc also has a prominent and short-lived C\,{\sc ii} $\lambda6580$ line that is even stronger than Si\,{\sc ii} $\lambda6355$ at $t \approx -15$~d and disappears at $t\approx -7$~d relative to $B$-band maximum, indicating a more complicated diversity among 91T/99aa-like objects. 

\citet{2018ApJ...852..100M} find that the pulsational delayed-detonation (PDDEL) models of \citet{2014MNRAS.441..532D} provide a good match to the early carbon features but a bad match to the ``bump''  in the light curve of iPTF16abc. In comparison with a typical SN~Ia (e.g., SN~2011fe; \citealt{2016AJ....151..125Z}),  spectroscopic and photometric observations show a very blue spectral energy distribution (SED) for iPTF16abc at very early times. The collision of SN ejecta with a nondegenerate companion \citep{2010ApJ...708.1025K} can produce an early blue bump, but this scenario is excluded for iPTF16abc by \citet{2018ApJ...852..100M} based on the early light curves. Other possible scenarios, including $^{56}$Ni mixing such that its distribution does not decrease monotonically toward the outer ejecta (e.g., \citealt{2019ApJ...870...13S, 2020A&A...642A.189M}) and ejecta interaction with circumstellar matter (CSM; e.g., \citealt{2023MNRAS.525..246H,2023MNRAS.521.1897M,2023MNRAS.522.6035M}), need further exploration. The reappearance of C\,{\sc ii} $\lambda6580$ after $B$-band maximum is also an interesting characteristic of iPTF16abc \citep{2024MNRAS.529.3838A}, which is also seen in SN~2018oh (a normal SN~Ia with an early flux excess; \citealt{2019ApJ...870...12L}). This reappearing carbon distinguishes iPTF16abc from 03fg-like or 02es-like objects, which have strong and long-lasting carbon features (e.g., 03fg-like SN~2020esm, \citealp{2022ApJ...927...78D}; 02es-like iPTF14atg,  \citealp{2015Natur.521..328C}). 

Some slow decliners among normal SNe~Ia, such as SN~2012cg \citep{2012ApJ...756L...7S,2016ApJ...820...92M} and SN 2013dy \citep{2013ApJ...778L..15Z,2015MNRAS.452.4307P}, share a few properties with the 91T/99aa-like subclass, including low velocity gradient and shallow silicon lines. Their spectra exhibit  intermediate-mass elements (IMEs; e.g., Si\,{\sc ii} $\lambda6355$) with moderate line strengths lying between those of 91T/99aa-like and other normal SNe~Ia. In particular, strong C\,{\sc ii} $\lambda6580$ absorption, comparable to the Si\,{\sc ii} $\lambda6355$ line, is seen in the early-time spectra of both SN~2012cg and SN 2013dy, while it is usually weak or absent in normal SNe~Ia (e.g., SN~2011fe; \citealt{2011Natur.480..344N,2016ApJ...820...67Z}). Combining this with the reappearing carbon feature seen in SN~2018oh, the 91T/99aa-like objects and normal SNe~Ia may share the peculiar  evolution of carbon absorption and thus a similar progenitor system and explosion mechanism. More 91T/99aa-like SNe~Ia with very early observations are needed to study their origins and their potential connections with normal SNe~Ia. 

In this paper, we present extensive photometry and spectroscopy of SN~2023wrk, a 91T/99aa-like SN~Ia. The observations are described in Section\,\ref{sec:obs}. We analyze the light curves and spectra in Sections\,\ref{sec:LC_ana} and \ref{sec:spec_ana}, respectively. A discussion is given in Section\,\ref{sec:discussion}, and we summarize in Section\,\ref{sec:conclusion}.

\section{OBSERVATIONS AND DATA REDUCTION} \label{sec:obs}

\begin{figure*}
    \centering
    \includegraphics[width=0.65\textwidth]{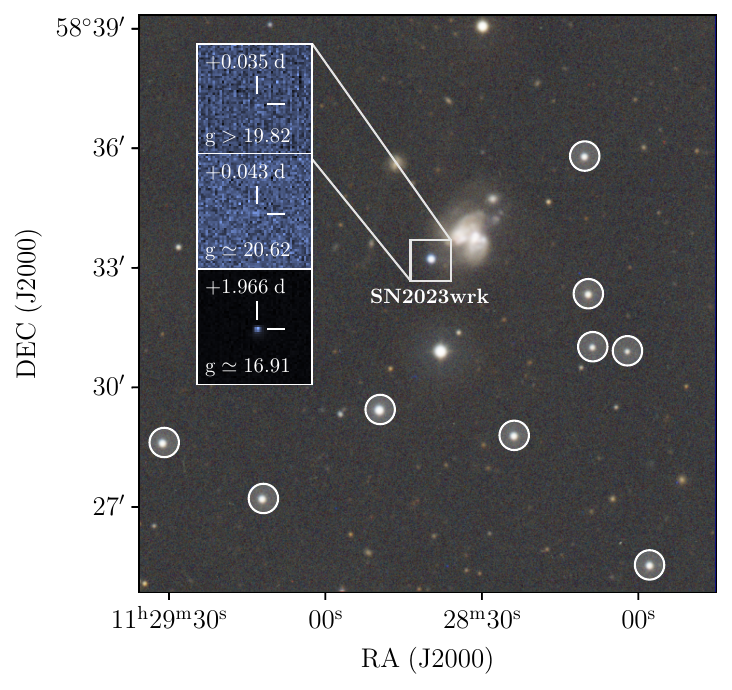}
    \caption{Schmidt 67/91 cm telescope $14.5' \times 14.5'$ color composite ($B/V/r$) image of SN~2023wrk and its host galaxy. The reference stars used to calibrate the photometry are marked with circles. The insets show the zoomed-in region of three ZTF difference images from IRSA, centered on the SN location, taken at phases of $t \approx +0.035$, +0.043, and +1.966 d relative to the time of first light (MJD 60251.48; see Section\,\ref{subsec:first_light}). The top inset gives a $5\sigma$ limit. 
    }
    \label{fig:position}
\end{figure*}

SN~2023wrk was discovered \citep{2023TNSTR2849....1G} on 2023 Nov. 4 (UTC dates are used throughout this paper; MJD 60252.233) by the Gravitational-wave Optical Transient Observer (GOTO) at $\alpha=11^{\rm hr}28^{\rm m}39.166^{\rm s}$, $\delta=58^\circ33^{\prime}12.68^{\prime\prime}$ (J2000), just $\sim 0.7$~d after the last nondetection by the Zwicky Transient Facility (ZTF) on MJD 60251.516. Note that ZTF has a $3\sigma$ detection on MJD 60251.523, as shown in Figure\,\ref{fig:position}, where the ZTF difference images in the insets are taken from IRSA \citep{https://doi.org/10.26131/irsa539}. The object was then classified as an SN~Ia based on a spectrum taken with YFOSC \citep{2019RAA....19..149W} on the Lijiang 2.4~m telescope (LJT; \citealt{2015RAA....15..918F}) of Yunnan Astronomical Observatories the day after discovery \citep{2023TNSCR2870....1L}. The prominent feature at $\sim 6250$~\AA\ comes from C\,{\sc ii} $\lambda6580$ instead of Si\,{\sc ii} $\lambda6355$, which may have led to the wrong estimated redshift by \citet{2023TNSCR2870....1L}. The interstellar Na\,{\sc i}\,D absorption shows that the merging binary galaxy NGC 3690 is the host of SN~2023wrk (see Section\,\ref{subsec:host}). 
As noted in \citet{2023TNSAN.296....1A}, SN~2023wrk is the first thermonuclear SN explosion in this extremely productive galaxy which has produced more than 15 SNe over the past 30 years.

\subsection{Photometry}\label{subsec:obs_phot}

After the discovery, we performed timely photometry of SN~2023wrk in $grizBVRI$ filters with the LJT, the Nanshan One-meter Wide-field Telescope (NOWT; \citealt{2020RAA....20..211B}) at Nanshan Station of Xinjiang Astronomical Observatory, the 0.8~m Tsinghua University-NAOC telescope (TNT; \citealt{2008ApJ...675..626W,2012RAA....12.1585H}) at Xinglong Station of NAOC, and the Schmidt 67/91 cm Telescope (67/91-ST) and the 1.82~m Copernico Telescope (Copernico) at the Mount Ekar Observatory in Italy.

We also triggered the Global Rapid Advanced Network Devoted to the Multi-messenger Addict (GRANDMA) to observe SN~2023wrk since 2023 Nov. 6. GRANDMA is a network of ground-based facilities dedicated to time-domain astronomy focused on electromagnetic follow-up observations of gravitational-wave progenitors and other transients. The network contains 36 telescopes from 30 observatories, 42 institutions, and groups from 20 countries. GRANDMA utilizes wide and narrow field-of-view telescopes distributed across all continents (\citealt{GRANDMAO3A,GRANDMA03B,2022MNRAS.515.6007A,2023ApJ...948L..12K}). GRANDMA's network extends to its citizen science project entitled Kilonova-Catcher (KNC)\footnote{\url{http://kilonovacatcher.in2p3.fr/}}, allowing amateur astronomers to contribute to  GRANDMA campaigns. 

All of these images were preprocessed following standard routines, including bias subtraction, flat-field correction, dark-current correction, and cosmic-ray removal. We have used \texttt{AUTOPHOT} \citep{2022A&A...667A..62B} to perform aperture or point-spread-function (PSF) photometry for the LJT, NOWT, TNT, 67/91-ST, and Copernico images. Aperture photometry was conducted for GRANDMA images using the Python package \texttt{STDPipe}\footnote{\url{https://gitlab.in2p3.fr/icare/stdpipe}} \citep{stdpipe}. The instrumental magnitudes were calibrated using the Gaia Synthetic Photometry Catalogue \citep{2023A&A...674A..33G} for $UBVRI$ filters, the Pan-STARRS Release 1 \citep{2016arXiv161205560C} for $griz$ filters, and the Sloan Digital Sky Survey Release 16 \citep{2020ApJS..249....3A} for the $u$ filter. In addition, the unfiltered images of KNC-Unistellar span a wide band from near-ultraviolet (near-UV) to near-infrared (NIR), posing challenges for standard calibration. We addressed these images by using the $G$ band of the Gaia eDR3 catalog \citep{2023A&A...674A..33G} because it covers a range of wavelengths similar to that of the Unistellar instruments (visible through NIR). This resulted in a color-term correction below 0.1 mag, rendering it negligible. 
% The $G$ band of Gaia covers the visible and near-infrared (NIR) domain, leading to the adoption of this calibration approach for the entire Unistellar dataset. 

SN~2023wrk was also observed by the Ultraviolet/Optical Telescope (UVOT; \citealp{2004ApJ...611.1005G,2005SSRv..120...95R}) on the {\it Neil Gehrels Swift Observatory} \citep{2004ApJ...611.1005G} in three UV ($UVW2$, $UVM2$, $UVW1$) and three optical ($U$, $B$, $V$) filters. We extracted {\it Swift} photometry using \texttt{HEASOFT}\footnote{HEASOFT, the High Energy Astrophysics Software, \url{https://www.swift.ac.uk/analysis/software.php}} with the latest \emph{Swift} calibration database\footnote{\url{https://heasarc.gsfc.nasa.gov/docs/heasarc/caldb/swift/}}. In addition, we include the $g$- and $r$-band photometry of ZTF provided by Lasair\footnote{https://lasair-ztf.lsst.ac.uk}. 

All the light curves of SN~2023wrk are shown in Figure\,\ref{fig:LC_all}. A summary of the photometric observations triggered by us is presented in Appendix\,\ref{sec:all_LCs}, including telescope names and the corresponding filters and number of images. The photometry is also presented in Appendix\,\ref{sec:all_LCs}.

\begin{figure*}
    \centering
    \includegraphics[width=0.7\textwidth]{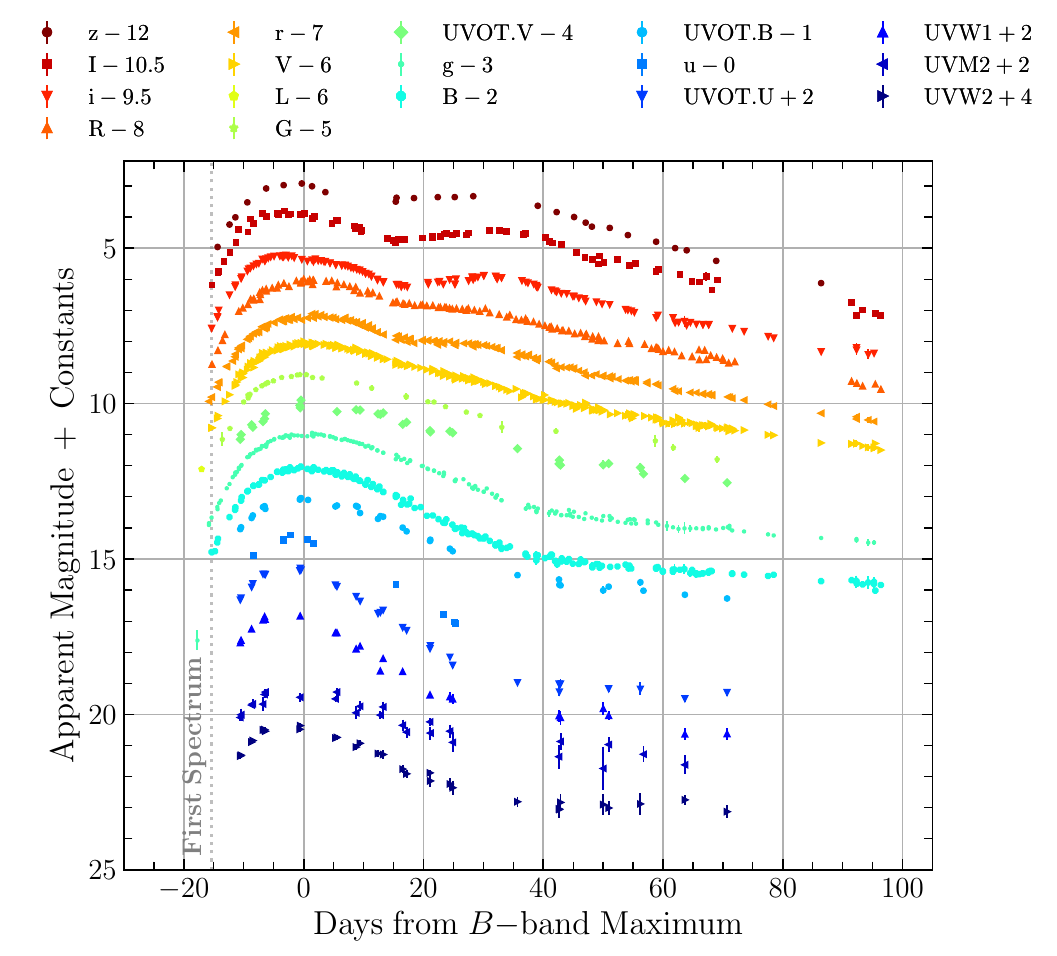}
    \caption{UV and optical light curves of SN~2023wrk. The phase is relative to the $B$-band maximum (MJD = 60269.45, see Section\,\ref{subsec:LC_op}). Data in different filters are shown with different colors and shapes, and are shifted vertically for better display, as indicated in the top legend. No S-correction (e.g., see \citealt{2002AJ....124.2100S}) has been applied to the data points plotted in this figure.
    }
    \label{fig:LC_all}
\end{figure*}

\subsection{Spectroscopy}\label{subsec:obs_spec}

\begin{figure*}
    \centering
    \includegraphics[width=0.85\textwidth]{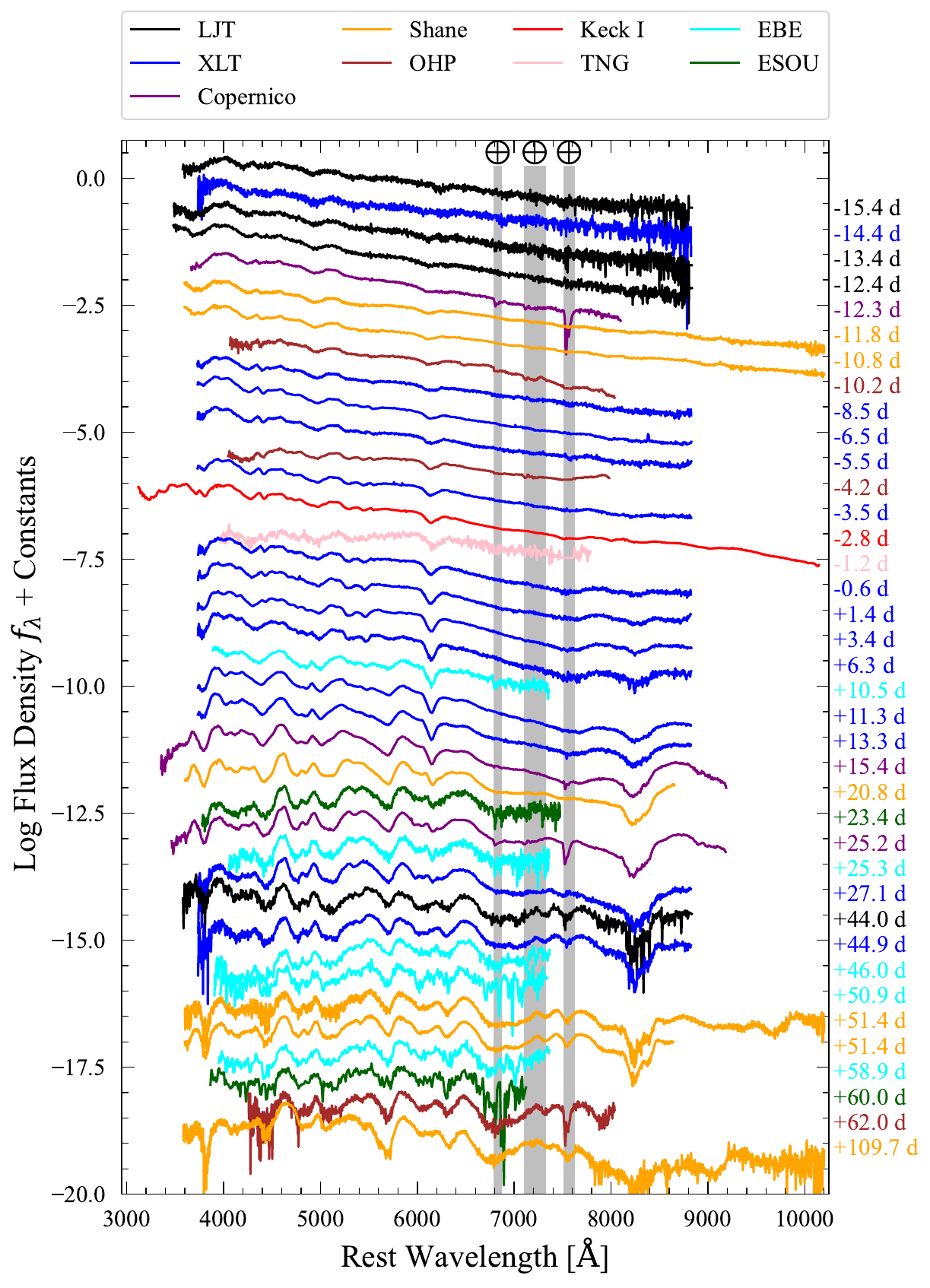}
    \caption{Optical spectral evolution of SN~2023wrk from $-15.4$ to +109.7 d relative to $B$-band maximum light. All spectra have been corrected for reddening and host-galaxy redshift. Spectra taken with different telescopes are in different colors as indicated in the top legend. The phase of each spectrum is shown on the right side. Regions of the main telluric absorption are marked by gray vertical bands. 
    %The telluric absorption is not removed from the spectra at $t=$-12.3 and +25.2 d.
    }
    \label{fig:spec_all}
\end{figure*}

Optical spectra of SN~2023wrk were collected with several different instruments, including BFOSC mounted on the Xinglong 2.16~m telescope \citep[XLT;][]{2016PASP..128j5004Z},  YFOSC on LJT, AFOSC on Copernico, LRIS \citep{1995PASP..107..375O} on the 10~m Keck-I telescope on Maunakea, Kast double spectrograph \citep{miller1993lick} and HIRES on the 3~m Shane telescope at Lick Observatory;  MISTRAL on the 1.93~m telescope at Observatoire de Haute Provence (OHP), DOLoRES-LRS on Telescopio Nazionale Galileo (TNG) on the island of La Palma, and two amateur telescopes EBE and ESOU of KNC. The Keck~I/LRIS spectrum was reduced using the \texttt{LPipe} pipeline \citep{2019PASP..131h4503P}. The standard \texttt{IRAF}\footnote{{IRAF} is distributed by the National Optical Astronomy Observatories, which are operated by the Association of Universities for Research in Astronomy, Inc., under cooperative agreement with the National Science Foundation (NSF).} routines were used to reduce other spectra. Flux calibration of the spectra was performed with spectrophotometric standard stars observed on the same nights. Atmospheric extinction was corrected with the extinction curves of local observatories. 
We derived the telluric correction using the spectrophotometric standard star. A journal of spectroscopic observations of SN~2023wrk is presented in Appendix\,\ref{sec:all_spec}, and all the spectra are displayed in Figure\,\ref{fig:spec_all}. These spectra are available in the Weizmann Interactive Supernova Data Repository (WISeREP)\footnote{\url{https://www.wiserep.org/}}. 

\section{PHOTOMETRY}\label{sec:LC_ana}

\subsection{Reddening and Distance}\label{subsec:host}

%Interstellar Na\,{\sc i}\,D absorption at about 0.01 
% can be clearly 
% the nearby binary galaxy NGC 3690 
Interstellar Na\,{\sc i}\,D absorptions 
with a redshift $z \approx 0.01$ can be clearly seen and should be from the host binary galaxy NGC 3690 \citep[$z = 0.0104$;][]{https://doi.org/10.26132/ned1}. The Galactic reddening is estimated to be $E(B-V)_{\rm Gal}=0.014$ mag (\citet{2011ApJ...737..103S}). The equivalent width (EW) of Na\,{\sc i}\,D absorption due to the host galaxy can be used to estimate the host reddening using the empirical relation ${\rm log_{10}}(E(B-V)) = 1.17 \times EW(D_1 + D_2) - 1.85$ \citep{2012MNRAS.426.1465P}. Rather than deriving the host reddening directly from this empirical relation, we estimate it by comparing the EWs of the Na\,{\sc i}\,D absorption due to the Milky Way and the host galaxy. With the spectrum taken at $t \approx -4.2$ d, we measure EWs of $0.33 \pm 0.03$~\AA\ and $0.58 \pm 0.05$~\AA\ for the Milky Way and the host-galaxy components, respectively. Using the empirical relation of \citet{2012MNRAS.426.1465P} and assuming $E(B-V)_{\rm Gal}=0.014$ mag, we obtain a host-galaxy reddening of $E(B-V)_{\rm host}=0.028 \pm 0.020$ mag 
for which we add a $\sim$68\% error as suggested by \citet{2013ApJ...779...38P}. 
Note that the reddening directly derived from the empirical relation is larger but still within the uncertainty, with $E(B-V)_{Gal}=0.034 \pm 0.023$ mag and $E(B-V)_{host}=0.067 \pm 0.046$ mag.
We adopt a host reddening of 0.028 mag and a total reddening of 0.042 mag in our analysis. Assuming H$_0 = 73\pm5$ km s$^{-1}$ Mpc$^{-1}$ (the value is taken from \citet{2022ApJ...934L...7R} and the uncertainty considers the Hubble tension) and correcting for peculiar motions related to the Virgo cluster and Great Attractor \citep{2000ApJ...529..786M}, the distance to NGC 3690 is estimated to be $48.25\pm3.32$ Mpc, with the corresponding distance modulus $33.42\pm0.15$ mag. 

\subsection{Time of First Light}\label{subsec:first_light}

\begin{figure}
    \centering
    \includegraphics[width=0.45\textwidth]{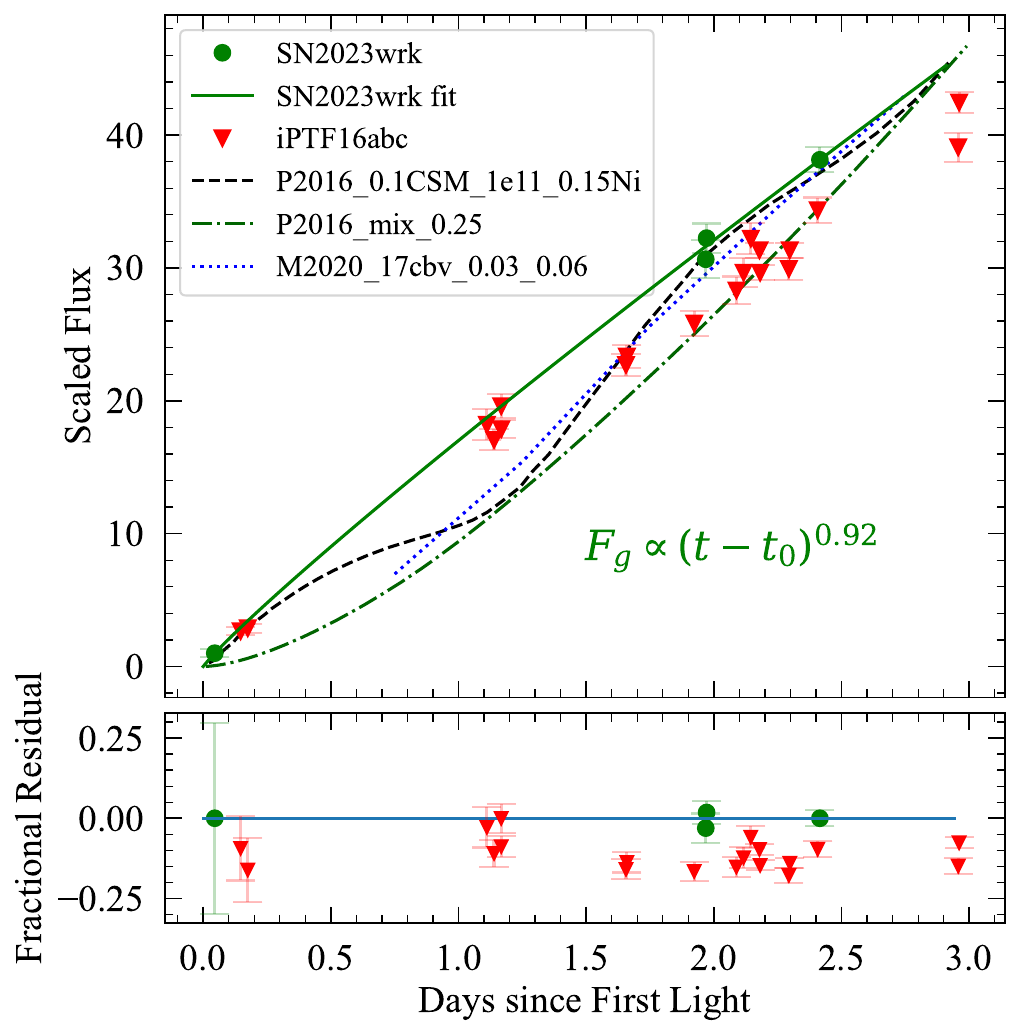}
    \includegraphics[width=0.45\textwidth]{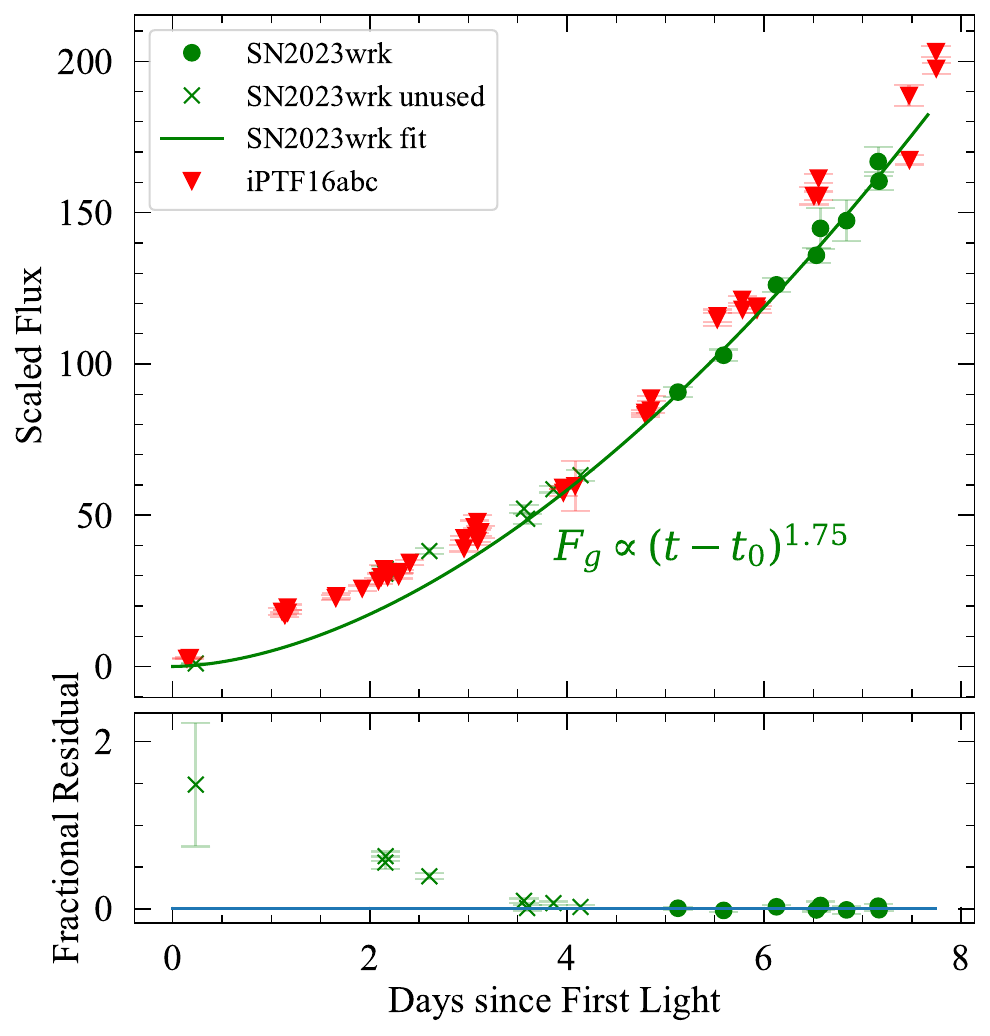}
    \caption{Power-law fits to the early-time $g$-band light curve of SN~2023wrk. Fits to the data obtained within the first 3 days and the first 7 days are shown in the top and bottom panels, respectively. The $g$  data of SN~2023wrk used to fit the power law are shown in green circles, with the unused data shown in green ``x'' symbols and the best-fit curve shown as a green line. For comparison, the $g$ data of iPTF16abc are shown as red inverted triangles. The fluxes of SN~2023wrk and iPTF16abc are converted from their absolute magnitudes and are then scaled to the first point of SN~2023wrk. The fractional residual is equal to $(F_{\rm obs}-F_{\rm fit})/F_{\rm fit}$, where $F_{\rm obs}$ is the flux of the observation and $F_{\rm fit}$ is the flux of the fitting curve.
    In the left panel, we plot the light curves of three models for comparison, and the corresponding parameters and discussions can be found in Section\,\ref{subsec:LC_model}.
    }
    \label{fig:LC_powerlaw}
\end{figure}

To estimate the time of first light, we fit a power law to our early data in $g$,
\begin{equation}
F_g=A(t-t_0)^\alpha ,
\label{eq:power_law}
\end{equation}
where $F_g$ is the flux density in the $g$ band, $A$ is the scale factor, $t_0$ is the time of first light, and $\alpha$ is the power-law index. Following \citet{2018ApJ...852..100M}, we fit the data within the first three days; this yields $t_0=60251.48\pm0.04$ (MJD) and $\alpha=0.92\pm0.17$. We fail to obtain the built-in samples using MCMC Hammer {\sc emcee} \citep{2013PASP..125..306F}, so the reported uncertainty is from the parameter covariances. The fitting results are shown in the top panel of Figure\,\ref{fig:LC_powerlaw}, where we add the $g$-band data of iPTF16abc for comparison. We find that the $g$ light curve of SN~2023wrk follows a power-law evolution with $\alpha \approx 1$, similar to iPTF16abc \citep{2018ApJ...852..100M}. This indicates that SN~2023wrk exhibits a prominent flux excess in the first few days, as does iPTF16abc. According to the fitting, the first ZTF observation of SN~2023wrk was taken only $\sim 0.05$ day after first light. We also apply a power-law fit using $g$-band data obtained within 4--7 days after first light, when the SN flux is $\lesssim 40$\% of the peak and the flux excess is negligible. The best fit is shown in the bottom panel of Figure\,\ref{fig:LC_powerlaw}, and it gives $t_0=60251.29$ and $\alpha=1.75$. In this case, the first point of SN~2023wrk is $\sim 0.1$ day later than that of iPTF16abc but with a lower flux. We find that {\sc emcee} gives $t_0=60251.09^{+0.31}_{-0.37}$ and $n=1.80^{+0.12}_{-0.10}$ based on the 16th, 50th, and 84th percentiles of the samples in the marginalized distributions. 
Such a first-light time derived from the linear fit to the first three days should be more accurate, since the epoch of the first detection is very close to that of the last nondetection. Moreover, a linear rise of early-phase flux has been observed in many SNe~Ia with early excess emission\citep{2018ApJ...865..149J}.
Thus, we adopt the first-light time as MJD 60251.48$\pm$0.04 in this paper. 

\subsection{Optical Light Curve}\label{subsec:LC_op}

We applied a 5th-order polynomial fit to the $B$- and $V$-band light curves around maximum light
and find that SN~2023wrk reached a peak magnitude of $B_{\rm max}=14.08\pm0.01$ mag on 2023 Nov. $21.45 \pm 0.10$ (MJD = $60269.45 \pm 0.10$) with $\Delta m_{15}(B)=0.87\pm0.02$ mag,  and $V_{\rm max}=14.08\pm0.01$ mag on 2023 Nov. $22.46 \pm 0.10$ (MJD = $60270.46 \pm 0.10$) with $\Delta m_{15}(V)=0.62\pm0.01$ mag; these are adopted throughout this paper. The rise time in $B$ is $t_{r}^{B}=17.79\pm0.11$ days, consistent with that of iPTF16abc ($t_{r}^{B}=17.91\pm^{0.07}_{0.15}$ days). Assuming $R_B=4.1$ and a total reddening of $E(B-V)=0.042$ mag, the $B$-band absolute peak magnitude is $M_{\rm max}(B)=-19.51\pm0.15$ mag. 

\begin{figure*}
    \centering
    \includegraphics[width=0.9\textwidth]{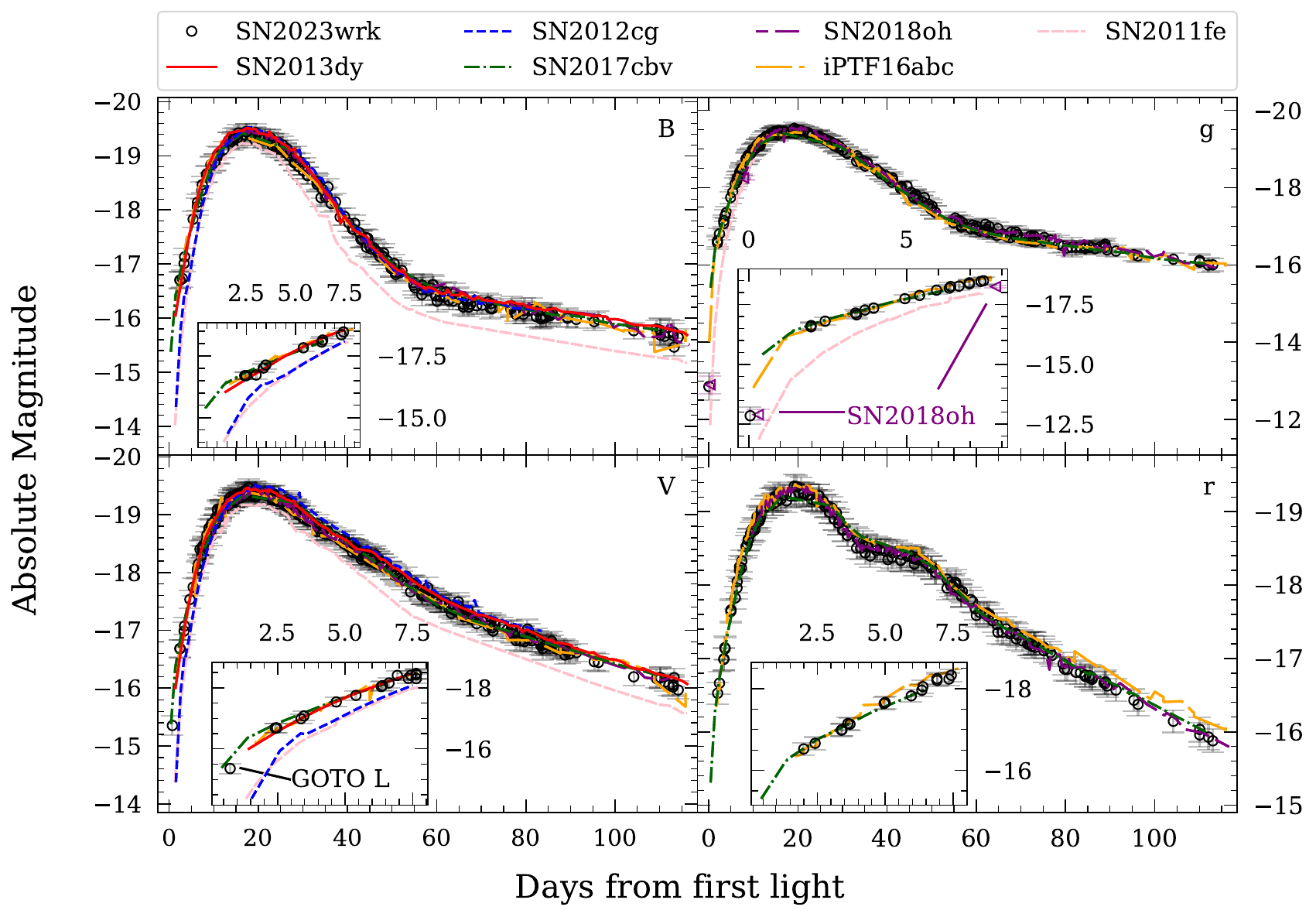}
    \caption{Light-curve comparison of SN~2023wrk with SNe~2011fe \citep{2011Natur.480..344N,2015MNRAS.446.3895F,2016ApJ...820...67Z}, 2012cg \citep{2019MNRAS.490.3882S}, 2013dy \citep{2013ApJ...778L..15Z,2015MNRAS.452.4307P}, 2017cbv \citep{2020ApJ...895..118B}, 2018oh \citep{2019ApJ...870...12L}, and iPTF16abc \citep{2018ApJ...852..100M}.  
    The first $L$-band detection by GOTO is plotted in the $V$-band subplot, as these two bands have a similar effective wavelength. The light curve of each object has been corrected for the reddening reported in the corresponding reference. 
    }
    \label{fig:LC_op_compare}
\end{figure*}

In Figure\,\ref{fig:LC_op_compare}, we compare the absolute $B$, $g$, $V$, and $r$ light curves of SN~2023wrk and some well-observed SNe~Ia, including the typical normal SN~2011fe ($\Delta m_{15}(B)=1.18\pm0.03$ mag; \citealt{2016ApJ...820...67Z}), the 91T/99aa-like object iPTF16abc ($\Delta m_{15}(B)=0.89\pm0.01$ mag\footnote{This value is obtained by applying a 5th-order polynomial fit to the $B$-band light curve as no $\Delta m_{15}(B)$ value is reported by \citet{2018ApJ...852..100M}.}, and four normal SNe~Ia which have been reported to have early flux excesses: SNe~2012cg ($\Delta m_{15}(B)=0.91\pm0.03$ mag; \citealt{2019MNRAS.490.3882S}), 2013dy ($\Delta m_{15}(B)=0.87\pm0.02$ mag; \citealt{2019MNRAS.490.3882S}), 2017cbv ($\Delta m_{15}(B)=0.99\pm0.01$ mag; \citealt{2020ApJ...904...14W}), and 2018oh ($\Delta m_{15}(B)=0.96\pm0.03$ mag; \citealt{2019ApJ...870...12L}). 
The adopted distance modulus, reddening (assuming $R_V=3.1$), and first-light time of the comparison SNe~Ia can be found in Appendix\,\ref{sec:other_para}. We find that SN~2023wrk has light curves similar to iPTF16abc and the normal SNe~Ia with early excesses except for SN~2012cg.  In the early phases, SN~2012cg appears obviously fainter than the other four comparison SNe~Ia (including SN~2023wrk) having an early excess, though they exhibit similar light curves at around the maximum light. In fact, SN~2012cg only exhibits weak early excesses compared to SN~2011fe. Beyond maximum light, SN~2023wrk, iPTF16abc, and all the normal SNe~Ia with early excesses have similar light-curve decline rates which are obviously slower than those of SN~2011fe.

\begin{figure*}
    \centering
    \includegraphics[width=0.7\textwidth]{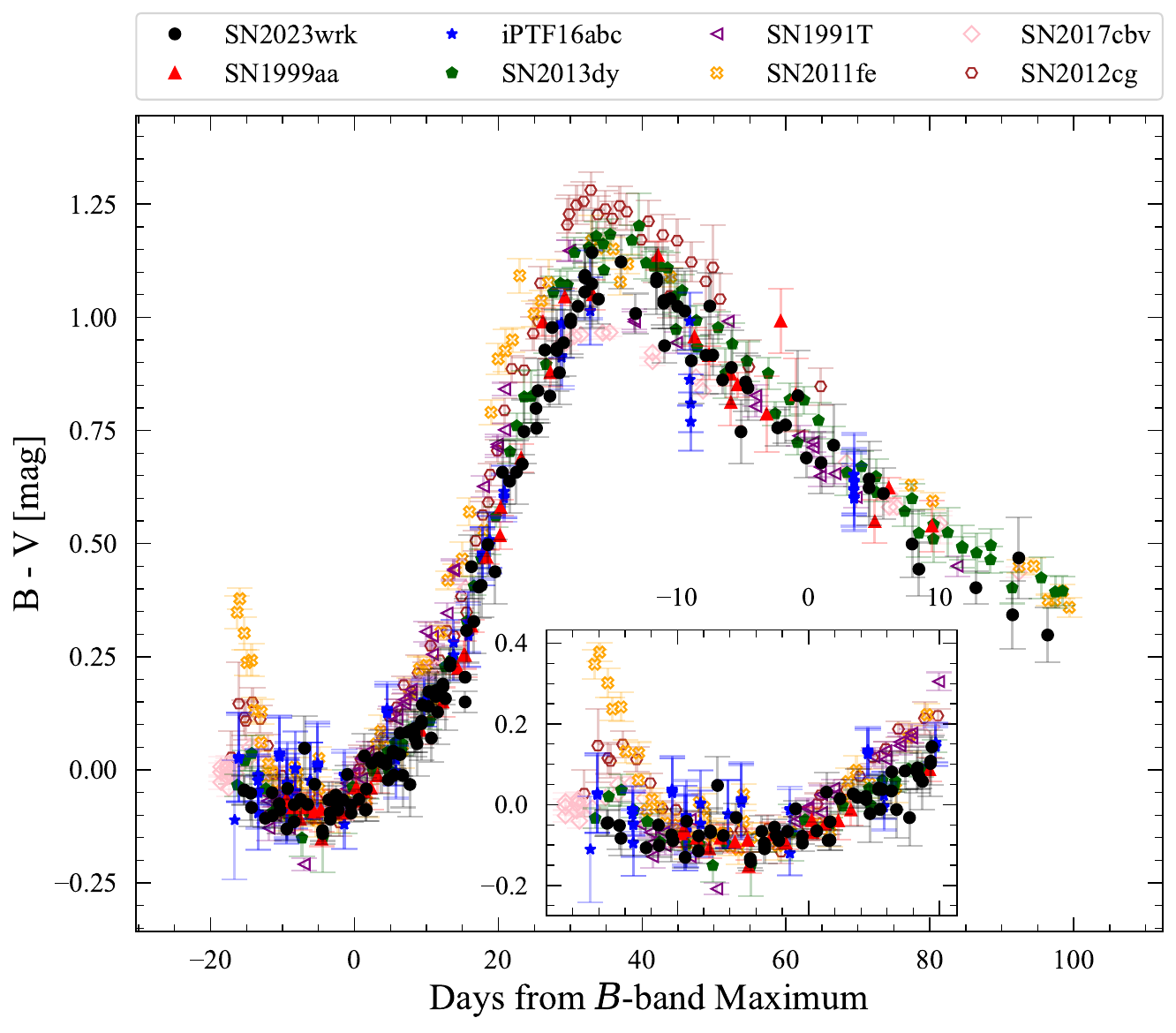}
    \caption{The $B-V$ color comparison between SN~2023wrk and SNe~1991T \citep{1998AJ....115..234L, 2004MNRAS.349.1344A}, 1999aa \citep{2006AJ....131..527J, 2008ApJ...686..749K}, 2011fe \citep{2016ApJ...820...67Z}, 2012cg \citep{2019MNRAS.490.3882S}, 2013dy \citep{2019MNRAS.490.3882S}, 2017cbv \citep{2020ApJ...895..118B}, and iPTF16abc \citep{2018ApJ...852..100M}. All of the color curves have been corrected for reddening due to the Milky Way and host galaxies. The inset shows the evolution during a phase from $-20$ to 10 days since $B$-band maximum light.}
    \label{fig:color_op_compare}
\end{figure*}

The $B-V$ color evolution is shown in Figure\,\ref{fig:color_op_compare}, where SN~2023wrk and iPTF16abc (and perhaps SN 1999aa) show a constant blue color until  maximum light. Among the comparison sample, SN 2011fe shows a rapid red-blue color evolution at $t\lesssim -10$ days relative to the $B$ maximum, while SNe~1991T, 2012cg, 2013dy, and 2017cbv tend to show a gradual red-to-blue color evolution within the first week after explosion. The overall color evolution of SN 2023wrk is quite similar to that of iPTF16abc, indicating that these two 99aa-like SNe~Ia share similar temperature evolution. 
 
\section{SPECTRAL ANALYSIS}\label{sec:spec_ana}

\subsection{Spectral Evolution}\label{subsec:spec_evo}

\begin{figure*}
    \centering
    \includegraphics[width=0.95\textwidth]{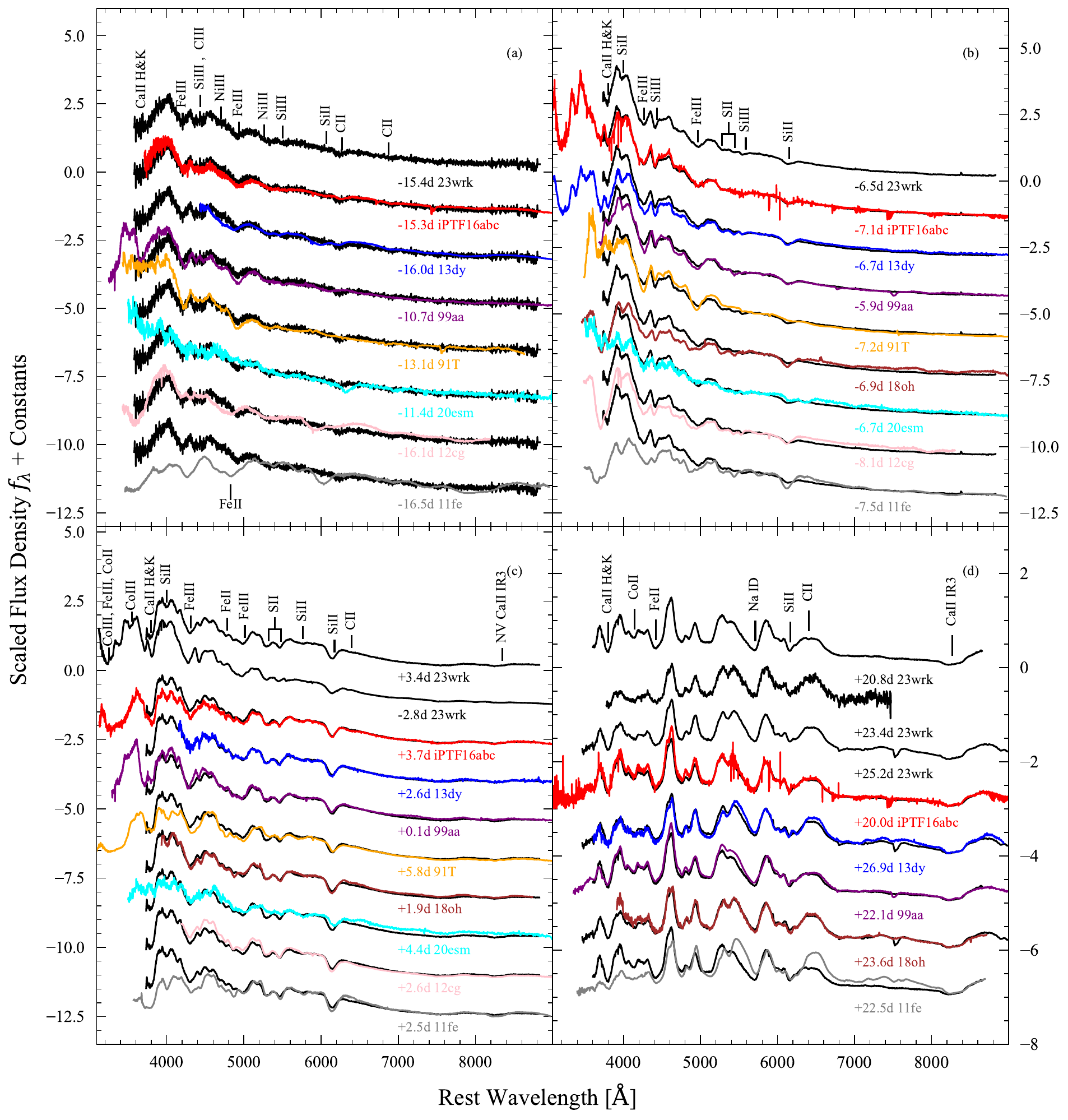}
    \caption{Spectral comparison of SN~2023wrk and SNe~1991T \citep{1992ApJ...384L..15F, 1995A&A...297..509M}, 1999aa \citep{2004AJ....128..387G, 2008AJ....135.1598M, 2008ApJ...686..749K, 2012MNRAS.425.1789S}, 2011fe \citep{2016ApJ...820...67Z, 2014MNRAS.439.1959M}, 2012cg \citep{2012ApJ...756L...7S,2016ApJ...820...92M,2020MNRAS.492.4325S}, 2013dy \citep{2013ApJ...778L..15Z, 2015MNRAS.452.4307P, 2016AJ....151..125Z}, 2018oh \citep{2019ApJ...870...12L,2023TNSAN.112....1Y}, 2020esm \citep{2022ApJ...927...78D}, and iPTF16abc \citep{2018ApJ...852..100M} at several selected epochs ($t \approx -15$, $-7$, +3, and +20 d relative to $B$-band maximum). The spectra have been corrected for reddening and host redshift.
    }
    \label{fig:spec_compare}
\end{figure*}

We compare the spectra of SN~2023wrk with those of comparison SNe~Ia in Figure \ref{fig:spec_compare}. At $t \approx -15$ d relative to $B$-band maximum, SN~2023wrk shows prominent Fe\,{\sc iii} absorption and weak Si\,{\sc ii} $\lambda6355$ absorption characteristic of 91T/99aa-like objects, though its Fe\,{\sc iii} lines are weaker. And it seems that SN~2023wrk is more similar to SN~1999aa as Ca\,{\sc ii} H\&K lines can be clearly seen. In addition, we identify Ni\,{\sc iii} features in the spectrum of SN~2023wrk by comparing it with the $t \approx -10.7$ d spectrum of SN~1999aa\citep{2004AJ....128..387G}, which indicates nickel in the very outer layers. Different from both SNe~1991T and 1999aa, however, SN~2023wrk exhibits strong C\,{\sc ii} $\lambda6580$ absorption which is a characteristic of 03fg-like objects such as SN~2020esm. We notice that all of these features of SN~2023wrk are also found in iPTF16abc, and the SEDs of these two SNe~Ia are also extremely similar and much bluer than SN~2011fe at short wavelengths. The spectra of the normal SNe~Ia having flux excesses at early times (SNe~2012cg and 2013dy) are also similar to those of SN~2023wrk to some extent, such as the bluer SEDs and weaker Si\,{\sc ii} lines compared with SN~2011fe, and possible Fe\,{\sc iii} lines around 4200~\AA\ and 4900~\AA. But the difference is that their Si\,{\sc ii} $\lambda6355$ line is still stronger than that of SN~2023wrk, and their 4900~\AA\ absorption features have a more extended blue wing, which could be attributed to Fe\,{\sc ii} absorption and thus imply comparable line strengths of Fe\,{\sc ii} and Fe\,{\sc iii}.

At $t \approx -7$ d, the C\,{\sc ii} absorption disappeared in SN~2023wrk and iPTF16abc, while it is still prominent in SN~2020esm. The the Si\,{\sc ii} absorption of SN~2023wrk can be seen clearly and has a similar line-strength to not only iPTF16abc but also SN~1999aa and the normal Ia SN~2013dy, while SN~1991T still shows no evident Si\,{\sc ii} absorption. The profile of the 4900 \AA\ absorption feature of SN~2023wrk is still different from those of normal SNe~Ia with early excesses, especially SN~2018oh which has two visible absorption troughs.

At around the peak brightness, the C\,{\sc ii} absorption reappeared in SN~2023wrk and iPTF16abc. The spectra of SN2023wrk resemble those of SNe~2012cg, 2018oh, and even SN~2011fe though the SED is still different in short wavelength. The good wavelength coverage of the LRIS spectrum of SN~2023wrk at $t \approx -2.8$ d allows us to find an absorption feature in the blue side of Ca\,{\sc ii} H\&K. In contrast to SN~1999aa, this absorption is weaker than that of Ca\,{\sc ii} H\&K and can be just attributed to Co\,{\sc iii} and Si\,{\sc iii} rather than HV Ca\,{\sc ii} H\&K \citep{2024MNRAS.529.3838A}. We also find Co\,{\sc iii} and Co\,{\sc ii} lines at $\lesssim 4000$~\AA\ in the LRIS spectrum. The reappearing C\,{\sc ii} absorption of SN~2023wrk can still be seen at $t \approx +20.8$ d and finally disappeared at $t \approx +25.2$ d. 

\begin{figure}
    \centering
    \includegraphics[width=0.45\textwidth]{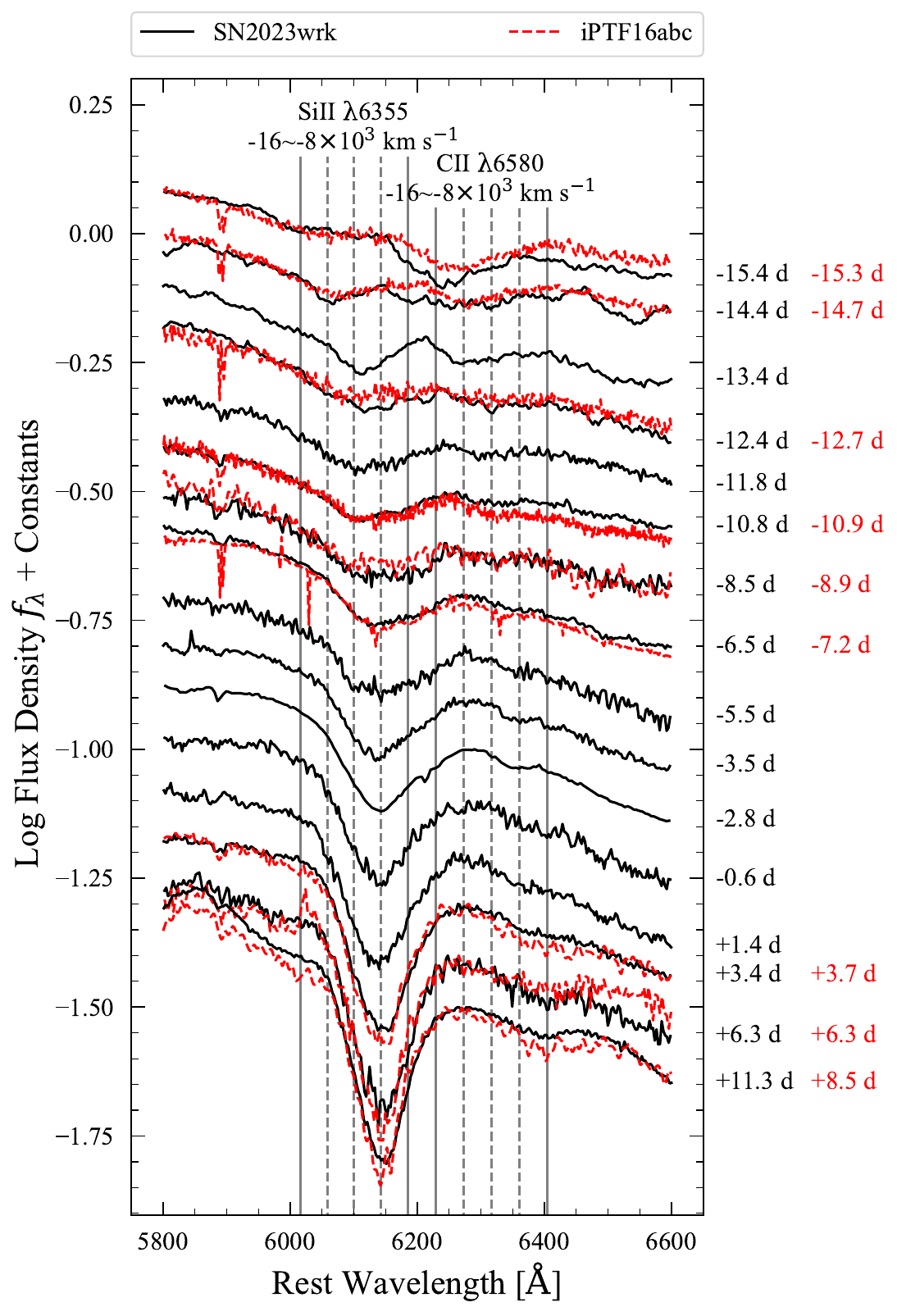}
    \caption{Evolution of the 5800--6600~\AA\ region in spectra of SN~2023wrk and iPTF16abc, spanning phases from $-15.4$ to +11.3 d relative to $B$ maximum. The spectra of SN~2023wrk at $t\approx -15.4$, $-14.4$, and $-13.4$ d are smoothed with a Savitsky-Golay filter of order 1 and a window of 40~\AA, and the spectrum at $t\approx -12.4$~d is smoothed with a window of 20~\AA. All  spectra have been corrected for reddening and host-galaxy redshift. The phase of each spectrum of SN~2023wrk is shown on the right side, and the phase of iPTF16abc is shown to the right of that of SN~2023wrk. The vertical lines mark the wavelength positions of Si\,{\sc ii} $\lambda6355$ and C\,{\sc ii} $\lambda6580$, corresponding to velocities from $-$16,000 to $-$8000 km s$^{-1}$ with an interval of 2000 km s$^{-1}$. Spectra of SN~2023wrk and iPTF16abc are shown in black solid lines and red dashed lines, respectively. 
    }
    \label{fig:Si_C_evo}
\end{figure}

One of the most interesting spectral features in SN~2023wrk is the evolution of the Si\,{\sc ii} $\lambda6355$ and C\,{\sc ii} $\lambda6580$ lines. To better examine this evolution, Figure\,\ref{fig:Si_C_evo} shows the 5800--6600~\AA\ region spanning phases from $-15.4$ to +11.3 d relative to $B$-band maximum brightness, overplotted with the spectra of iPTF16abc at comparable phases. At $t \approx -15.4$ d, the Si\,{\sc ii} absorption is nearly absent, while the C\,{\sc ii} line is prominent with a velocity of $\sim 15,500$ km s$^{-1}$.
One day later, the Si\,{\sc ii} line becomes prominent with a velocity of $\sim 14,000$ km s$^{-1}$, while the C\,{\sc ii} line becomes progressively weaker within a few days and tends to disappear by $t \approx -7$ d. Interestingly, the C\,{\sc ii} absorption becomes visible again in the $t\approx -3.7$ d and $-$2.9 d spectra, and it grows even stronger at $t\approx +6.3$ d. By $t\approx +25$ d, the C\,{\sc ii} line finally disappears in the spectra (cf. Figure\,\ref{fig:spec_compare}).    

In summary, the spectra of SN~2023wrk are overall quite similar to those of iPTF16abc, with characteristics of 91T/99aa-like objects and a complex evolution of the C\,{\sc ii} absorption. Together with the similar photometric properties, it is likely that SN~2023wrk and iPTF16abc can be regarded as twin objects.

\subsection{Line Velocities}\label{subsec:line_v}

To better measure the line velocities of SN~2023wrk, we first smooth the spectra with a Savitsky-Golay filter \citep{1964AnaCh..36.1627S} of order 1 and appropriate width\footnote{We increase the filter width gradually until the minimum of the feature can be clearly identified.} and then normalize the spectra with a local pseudocontinuum for each feature. The pseudocontinuum points connecting the red and blue sides of the features are chosen interactively. The ejecta velocities are measured from the local absorption minimum of the features. The associated uncertainties are estimated by the standard deviation of a Monte Carlo sample, where we randomly vary the endpoints of the pseudocontinuum within 10~\AA\ and the filter width within an appropriate range where the line feature is not overly smoothed. 

\begin{figure}
    \centering
    \includegraphics[width=0.45\textwidth]{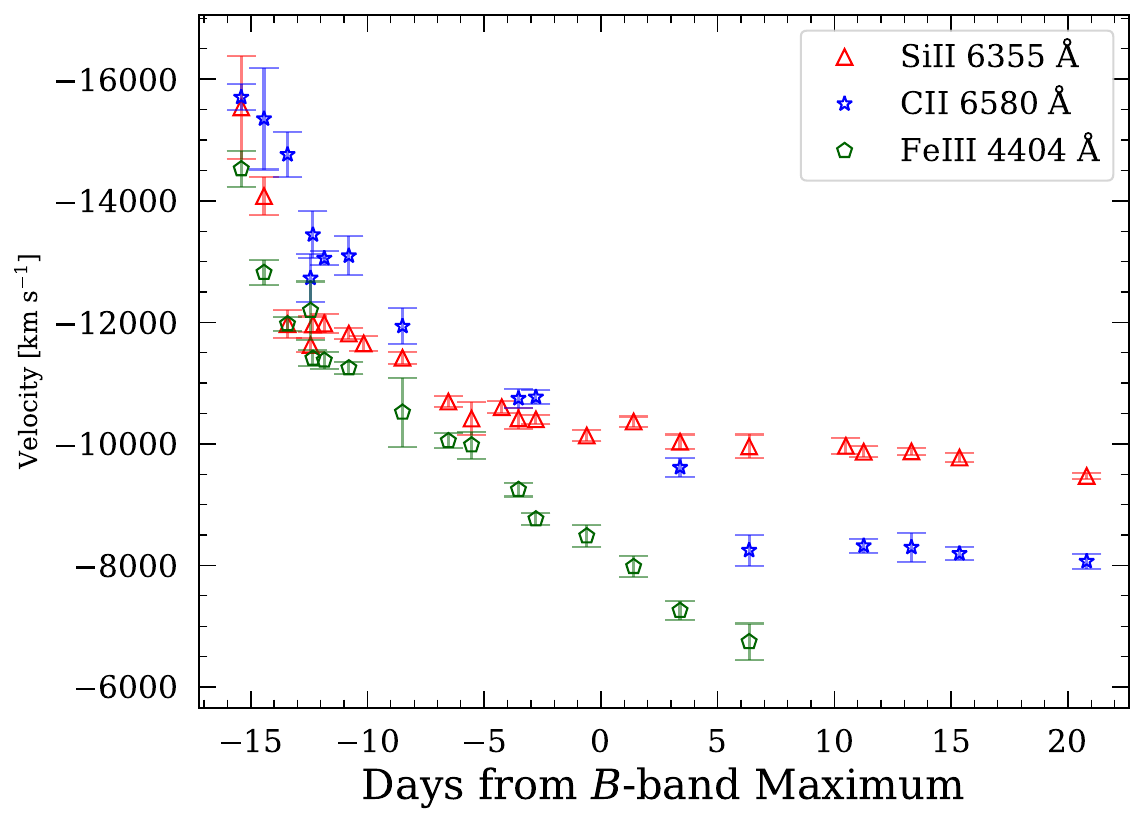}
    \includegraphics[width=0.45\textwidth]{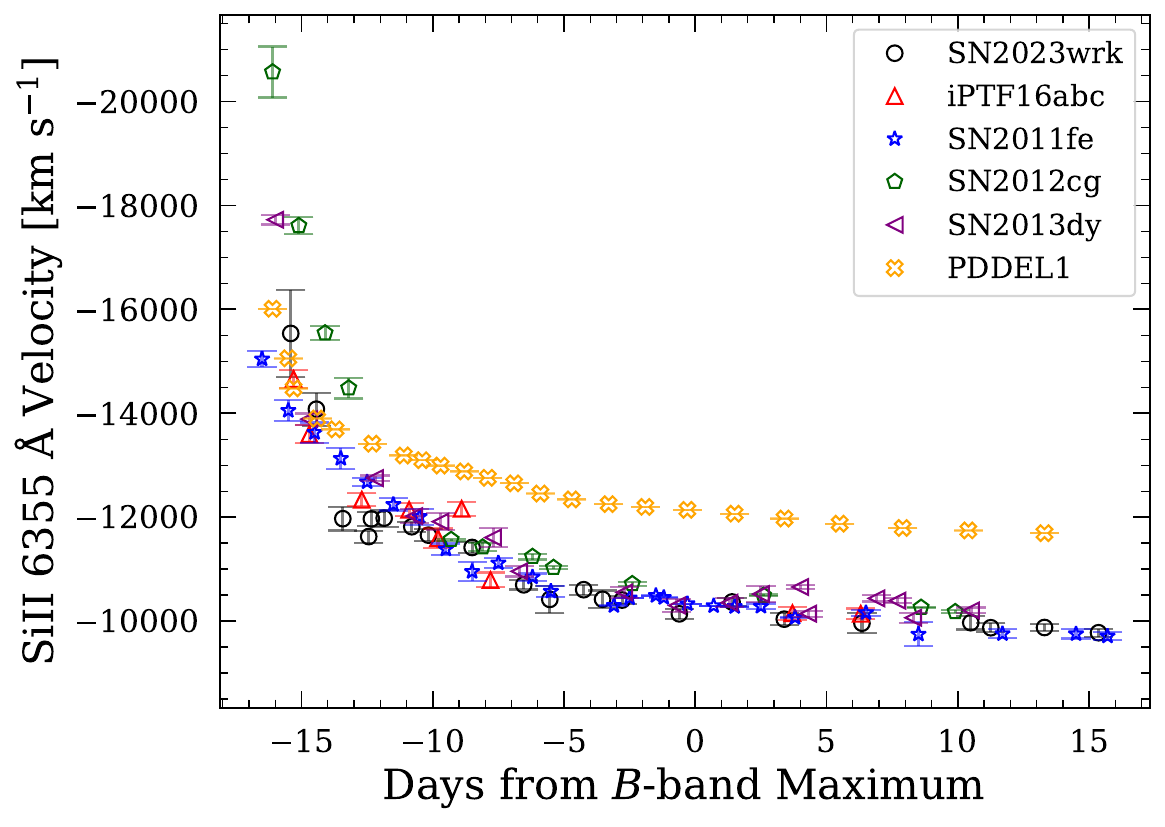}
    \caption{{\it Top:} Evolution of velocity inferred from absorption minima of Si\,{\sc ii} $\lambda6355$, C\,{\sc ii} $\lambda6580$, and Fe\,{\sc ii} $\lambda4404$ lines in SN~2023wrk. {\it Bottom:} Comparison of Si\,{\sc ii} $\lambda6355$ velocity evolution between SN~2023wrk and SNe~2011fe \citep{2016ApJ...820...67Z,2014MNRAS.439.1959M}, 2013dy \citep{2013ApJ...778L..15Z,2015MNRAS.452.4307P,2016AJ....151..125Z}, iPTF16abc \citep{2018ApJ...852..100M}, and the PDDEL1 model \citep{2014MNRAS.441..532D}. %Velocities of different lines/objects are shown in different colors and shapes as indicated in the top right legend. 
    }
    \label{fig:spec_velocities}
\end{figure}

For SN 2023wrk, the ejecta velocities inferred from absorption lines of different species are shown in the top panel of Figure \ref{fig:spec_velocities}. The Si\,{\sc ii} $\lambda6355$ velocity is measured as $10,100\pm100$ km s$^{-1}$ around the time of maximum light, and the velocity gradient is found to be $31 \pm 12$ km s$^{-1}$ day$^{-1}$ by fitting a linear function from $t\approx +0$ to $\sim +10$ d relative to $B$-band maximum. This puts SN~2023wrk in the NV \citep{2009ApJ...699L.139W} and LVG subgroups \citep{2005ApJ...623.1011B}. 
At $t\lesssim -10$~d, the C\,{\sc ii} $\lambda$6580 line has a velocity larger than that inferred from Si\,{\sc ii} $\lambda$6355, while at $t \gtrsim +5$ d the carbon velocity seems to experience a sudden drop, suggesting that the carbon in SN 2023wrk has a wider distribution than the silicon. 
In contrast with the nearly constant velocity of Si\,{\sc ii} after the peak light, the Fe\,{\sc iii} $\lambda4404$ line shows a rapid velocity decline, similar to that of iPTF16abc \citep{2024MNRAS.529.3838A} and other 91T/99aa-like objects \citep{2004AJ....128..387G}.
We summarize some basic photometric and spectroscopic parameters of SN~2023wrk in Table\,\ref{table:parameters}.

In the bottom panel of Figure \ref{fig:spec_velocities}, we show the Si\,{\sc ii} $\lambda6355$ velocity evolution for SN~2023wrk and the comparison sample. At $t\lesssim -14$~d, SNe~2012cg, 2013dy, iPTF16abc, and 2023wrk show larger velocities and also decline faster than SN 2011fe. 
The Si\,{\sc ii} velocity of SN~2023wrk and iPTF16abc shows a plateau-like evolution during  phases from $t\approx -13.5$ to $-8.5$~d relative to $B$-band maximum. A similar Si\,{\sc ii} velocity evolution may be also seen in SN~2013dy at $t\approx -10$~d. This velocity plateau may result from a spike structure in the density profile in velocity space and will be discussed in Section\,\ref{subsec:spec_model}.

\begin{table}[h]
\centering
\caption{Parameters of SN~2023wrk}\label{table:parameters}%
\begin{tabular}{ll}
Parameter & Value \\
\hline
\hline
$B_{\rm max}$ & $14.08 \pm 0.01$ mag\\
$M_{\rm max}(B)$ & $-19.51 \pm 0.15$ mag\\
$E(B-V)_{\rm host}$ & $0.028 \pm 0.020$ mag\\
$\Delta m_{15}(B)$ & $0.87 \pm 0.02$ mag\\
$t_{\rm max}(B)$ & $60269.45 \pm 0.10$ d\\
$t_0$ & $60251.48\pm0.04$ d\\
$t_{r}^{B}$ & $17.79\pm0.11$ d \\
$t_{r}^{bol}$ & $16.44\pm0.09$ d \\
$L_{\rm max}^{\rm bol}$ & $(1.71 \pm 0.24) \times 10^{43}$ erg s$^{-1}$\\
$M_{\rm ^{56}Ni}$ & $0.76 \pm 0.11$ M$_\odot$\\
$\mu$ & $33.42 \pm 0.15$ mag\\
$v_0({\rm Si\,\textsc{II}})$ & $10,100 \pm 100$ km s$^{-1}$\\
$\dot{v}({\rm Si\,\textsc{II}})$ & $31 \pm 12$ km s$^{-1}$ day$^{-1}$\\
%$R({\rm Si\,{\sc ii}})$ & \\
\hline
\end{tabular}
\end{table}

\section{DISCUSSION}\label{sec:discussion}

\subsection{Pseudobolometric Light Curve}\label{subsec:bolo}

We compute the pseudobolometric light curve of SN~2023wrk using {\it Swift} $UVW2$, $UVM2$, $UVW1$, $U$, and ground-based $BgVriz$ photometry. The data are interpolated in each filter so that the bolometric luminosity can be calculated at the same phase. Using \texttt{SNooPy2} \citep{2011AJ....141...19B}, the flux from $\sim 2000$ to 8800~\AA\ is estimated with the direct integrating method, and the flux from $\sim 8800$ to 24,000~\AA\ is estimated with the SED method based on the template spectra from \citet{2007ApJ...663.1187H}, since we do not have NIR data. To check whether our estimate of the NIR flux is reliable, we compare the flux density of the photometry, the template spectra mangled by the photometry of SN~2023wrk, the LRIS spectrum (3150--10,250~\AA) of SN~2023wrk, and a synthetic spectrum of the PDDEL1 model \citep{2014MNRAS.441..532D} at $t \approx -3$ d. 
We find that the $\sim 9000$--10,000~\AA\ region of the mangled template spectrum is close to that of SN~2023wrk, and the $\gtrsim 10,000$~\AA\ region of the mangled template spectrum is comparable to that of the PDDEL1 spectrum which agrees well with SN~2023wrk at $\gtrsim 5000$~\AA. These comparisons favor our estimation of the NIR flux of SN~2023wrk.
Then we obtain a pseudobolometric light curve from $\sim 2000$ to 24,000~\AA\, with phase coverage from $t\approx -10.6$ to +69.0 d relative to $B$-band maximum. The result is presented in Appendix\,\ref{sec:all_LCs}. 

\begin{figure*}
    \centering
    \includegraphics[width=0.7\textwidth]{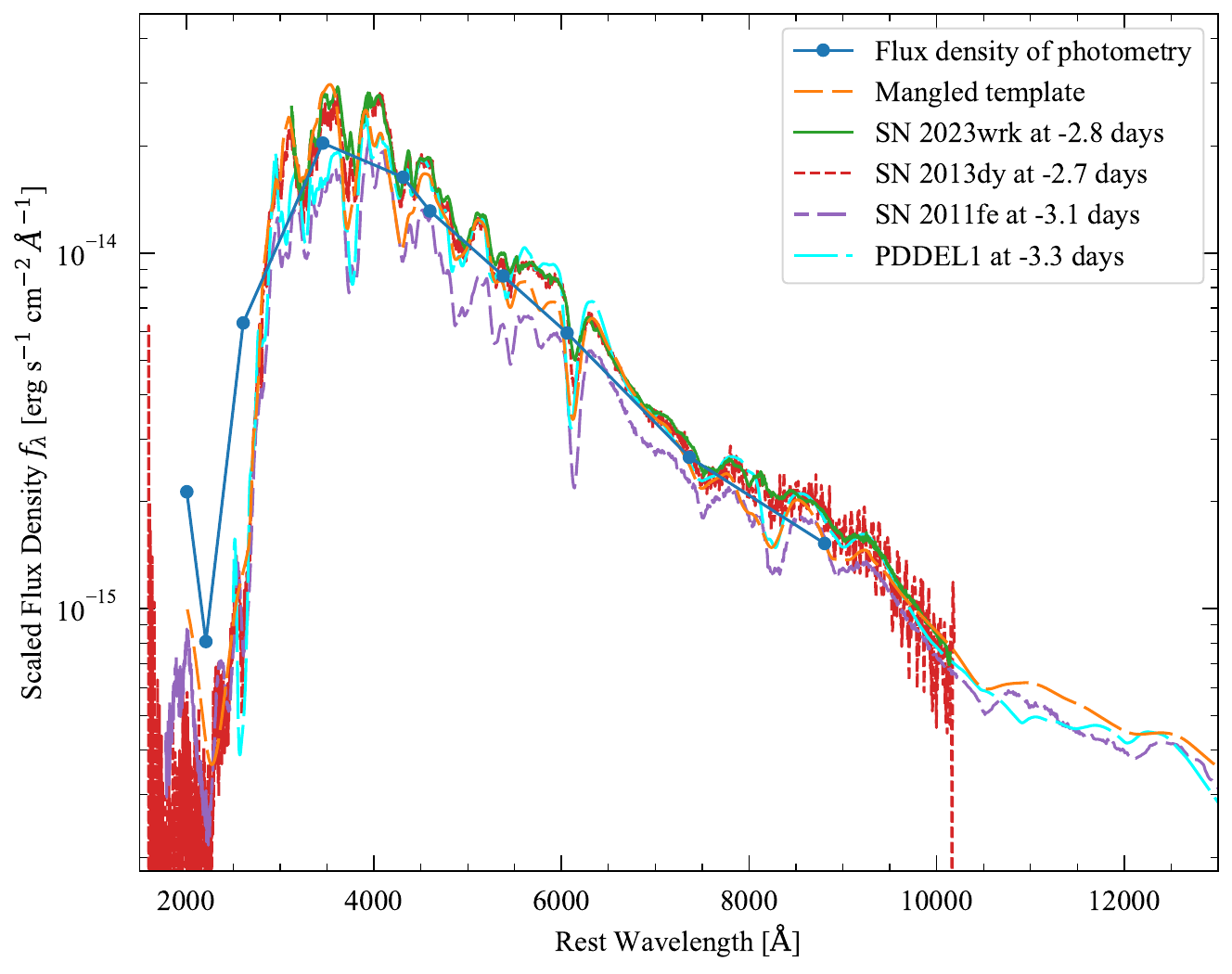}
    \caption{Comparison of the LRIS spectrum of SN~2023wrk and the synthetic spectrum of \citet{2007ApJ...663.1187H} mangled by the photometry of SN~2023wrk at $t\approx -3$ d. Spectra of SN~2011fe \citep{2014MNRAS.439.1959M} and SN 2013dy \citep{2015MNRAS.452.4307P}, and the synthetic spectrum of a PDDEL model (PDDEL1; \citealt{2014MNRAS.441..532D}) at similar epochs, are overplotted. All spectra have been corrected for reddening and host redshift. The flux density is scaled to the distance modulus of SN~2023wrk.
    }
    \label{fig:peak_SED}
\end{figure*}

We use the peak pseudobolometric luminosity to estimate the mass of $^{56}$Ni synthesized in the explosion following \citet{2005A&A...431..423S},
\begin{equation}
L_{\rm max}= (6.34e^{-\frac{t_{\rm r}}{8.8\ d}}+1.45e^{-\frac{t_{\rm r}}{111.3\ d}})\frac{M_{\rm Ni}}{{\rm M_\odot}}\times 10^{43}\,{\rm erg~s}^{-1},
\label{eq:MNi}
\end{equation}
where $t_{\rm r}$ is the rise time of the pseudobolometric light curve.
Using a 5th-order polynomial fit to the pseudobolometric light curve, we find a peak luminosity of $(1.71\pm0.24) \times 10^{43}$ erg s$^{-1}$ on MJD $60268.09\pm0.08$ and a rise time of $t_r^{\rm bolo}=16.44\pm0.09$ day. The pseudobolometric light curve peaks at about $-$1.35 d relative to the $B$-band maximum, consistent with the statistical result from \citet{2014MNRAS.440.1498S} who give a mean difference of about $-$1 d. Then we derive a $^{56}$Ni mass of $0.76\pm0.11$\,M$_{\odot}$, which is in line with the $^{56}$Ni mass of iPTF16abc (0.76\,M$_{\odot}$) deduced by \citet{2024MNRAS.529.3838A}. 
In addition, we notice that the spectrum of SN~2023wrk is comparable with that of SN~2013dy and evidently more luminous than that of SN~2011fe around maximum light as shown in Figure\,\ref{fig:peak_SED}.

\subsection{Comparison of SN~2023wrk and Models} \label{subsec:origin}

As presented and discussed in the previous sections, SN~2023wrk and iPTF16abc share extremely similar properties in both photometric and spectroscopic evolution, and can be considered twins of 99aa-like SNe~Ia. 
In addition, both of these SNe~Ia exploded at locations that are far from the center of their host galaxies, which may be related to galaxy mergers. The host of SN~2023wrk (NGC 3690) is a merging pair of galaxies, while the host of iPTF16abc (NGC 5221) has a long tidal tail that may arise from a galaxy merger. The above similarities suggest that these two SNe~Ia may have the same physical origin. 

\citet{2018ApJ...852..100M} performed comprehensive model comparisons for iPTF16abc. They found that the early-time light curve and color evolution of iPTF16abc can be best matched by the PDDEL models of \citet{2014MNRAS.441..532D} and the nickel mixing models of \citet{2016ApJ...826...96P}. The SN shock breakout and the companion interaction models are excluded by their early observations. They argued that either ejecta interaction with nearby unbound material or vigorous mixing of $^{56}$Ni to the outer region, or a combination of the two, can lead to the early-time characteristics of iPTF16abc. However, as \citet{2018ApJ...852..100M} noted, the PDDEL and nickel mixing models need improvements to interpret the observations better. 

\subsubsection{Light-Curve Comparison}\label{subsec:LC_model}

\begin{figure*}
    \centering
    \includegraphics[width=0.9\textwidth]{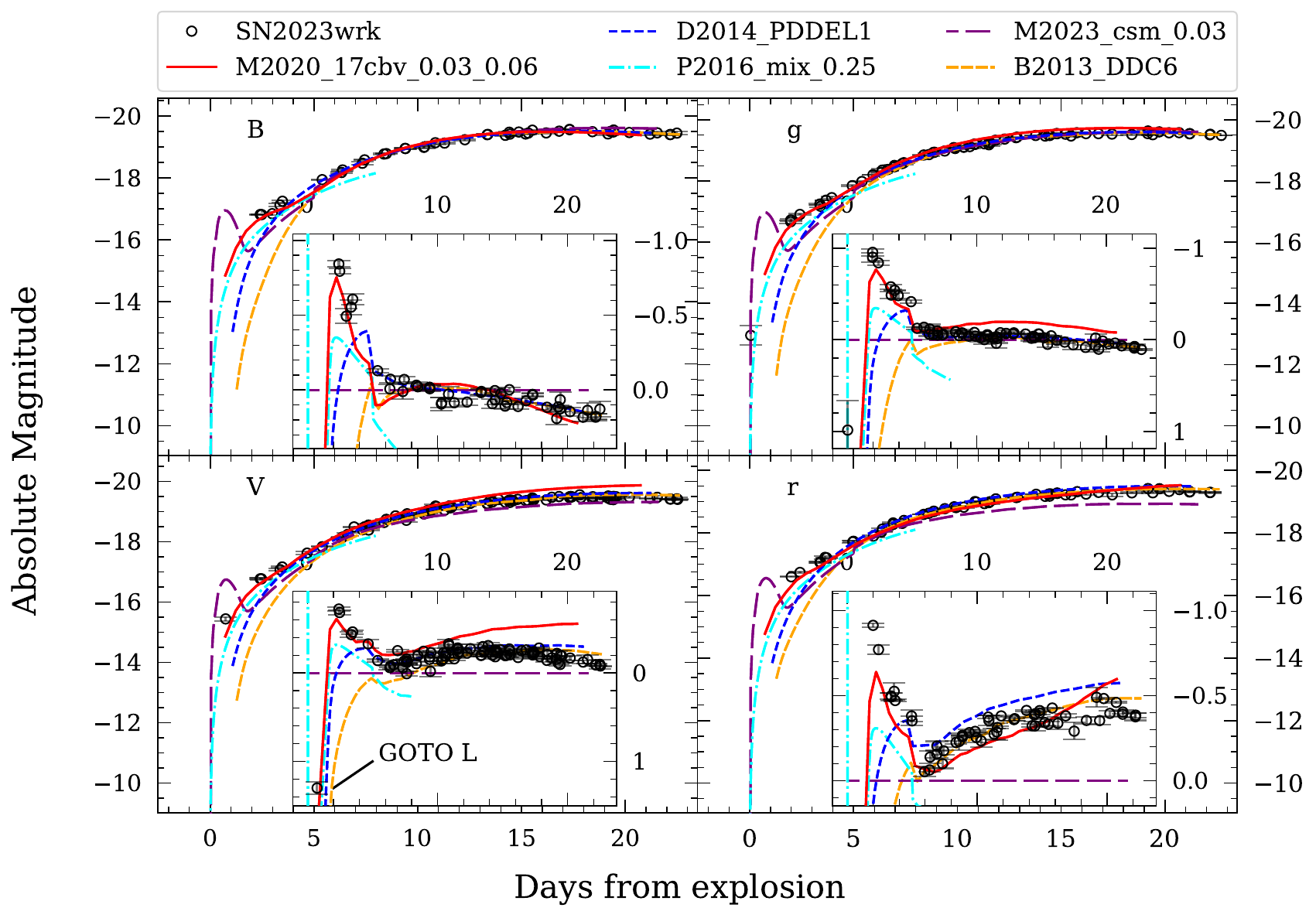}
    \caption{Comparison of early $BVgr$-band light curves between SN~2023wrk and some models, including an extended nickel mixing model of \citet{2016ApJ...826...96P}, which uses a boxcar averaging routine with width 0.25\,M$_{\odot}$ (P2016\_mix\_0.25), a standard delayed-detonation model of \citet{2013MNRAS.429.2127B} with $M_{\rm ^{56}Ni}=0.72\,{\rm M}_{\odot}$ (DDC6), a $^{56}$Ni=shell model of \citep{2020A&A...642A.189M}, which is based on their fiducial SN~2017cbv model and has a $^{56}$Ni shell of 0.03\,M$_{\odot}$ with a width of 0.06\,M$_{\odot}$ (M2020\_17cbv\_0.03\_0.06), a carbon-rich CSM interaction model of \citet{2023MNRAS.521.1897M} based on a super-Chandrasekhar mass (super-$M_{\rm Ch}$) explosion (M2023\_csm\_0.03), and a pulsational delayed-detonation model of \citet{2014MNRAS.441..532D} with $M_{\rm ^{56}Ni}=0.75\,{\rm M}_{\odot}$ (PDDEL1). The first detection by GOTO in the $L$ band is plotted in the $V$-band subplot. The light curve of SN~2023wrk has been corrected for reddening. The insets show the residuals relative to the M2023\_csm\_0.03 model. Note that the light curve of the M2023\_csm\_0.03 model is a combination of results from SuperNova Explosion Code ($t\lesssim +5$ d; with interaction power) and the direct convolution of spectra ($t\gtrsim +5$ d; more realistic when interaction power is negligible). 
    }
    \label{fig:model_lc}
\end{figure*}

Comparisons of the early light curve of SN~2023wrk and those predicted by some models are shown in Figure \ref{fig:model_lc}, where we assume that the time of first light and explosion are the same for SN~2023wrk, since this SN has nickel mixed in the outer layers. Note that a SN~Ia could experience a dark phase lasting for a few days if its $^{56}$Ni is confined to the innermost region \citep{2014ApJ...784...85P}. The $B$, $g$, and $V$ light curves of SN~2023wrk are matched best by a PDDEL model with $\Delta m_{15}(B)=0.95$~mag (PDDEL1) from $t\approx +4$ d after explosion to the peak, though the $r$-band light curve by PDDEL1 appears slightly brighter. The main defect is that PDDEL1 fails to match the luminosity bump of SN~2023wrk at $t\lesssim +4$ d. (i) The power-law index of the light curve of SN~2023wrk is broken at $t\approx +4$ d (see also Figure\,\ref{fig:LC_powerlaw}), which makes the light curve at $t\lesssim +4$ d look like a bump. However, such a transition in the light curve of PDDEL1 occurs at $t\approx +1$ d, indicating a much shorter duration for the bump than observed in SN~2023wrk. (ii) The PDDEL1 model is still too faint at $t\lesssim +4$ d, though it has extra radiation from the initial shock-deposited energy in the outer ejecta. \citet{2018ApJ...852..100M} mentioned that adding nickel mixing to the PDDEL models could reduce this inconsistency.

Here we consider two mixing modes: (i) the extended mixing of \citet{2016ApJ...826...96P} which has a mass-fraction profile of $^{56}$Ni monotonically decreasing toward the outer ejecta; and (ii) the $^{56}$Ni shell model of \citet{2020A&A...642A.189M} which has a high mass fraction of $^{56}$Ni in the outer layers. We notice that the extended nickel-mixing model of \citet{2016ApJ...826...96P} does not produce a bump, while the $^{56}$Ni-shell model of \citet{2020A&A...642A.189M} predicts a bump which breaks at $t\approx +4$ d after explosion. In fact, the light curve of SN~2023wrk matches best with the $^{56}$Ni-shell model at $t\lesssim +4$ d. 

The circumstellar matter (CSM) interaction model predicts a longer duration for the early excess emission compared with the PDDEL1 model, but also a peak within a few days after explosion which is not seen in SN~2023wrk. A combination of CSM interaction and  nickel mixing may improve the match. Specifically, the energy from the decay of the outward-mixing $^{56}$Ni could prevent the flux from declining after the CSM-interaction peak so that we would not see a peak within the first few days.
In Figure\,\ref{fig:LC_powerlaw}, we compare the first-three-days light curve of SN~2023wrk with an extended nickel mixing model P2016\_mix\_0.25, a $^{56}$Ni-shell model M2020\_17cbv\_0.03\_0.06, and a CSM + extended nickel mixing model P2016\_0.1CSM\_1e11\_0.15Ni which has a boxcar averaging routine with a width 0.25\,M$_{\odot}$ and 0.1\,M$_{\odot}$ of CSM located at a radius of 10$^{11}$ cm \citep{2016ApJ...826...96P}.
The $^{56}$Ni-shell model seems to have a nearly linear rise if we shift the time axis, but this shift means that the first-light time is $\sim 0.3$ days later than the explosion time, which conflicts with the observation of $^{56}$Ni in the outer layers. The scaled flux yielded from the extended nickel mixing model alone has a nonlinear rise, but that from the CSM + extended nickel mixing model matches well with the observations of SN~2023wrk, though the scaled flux at $t\approx +1$ d is lower than the near-linear fit of SN~2023wrk. Some adjustments to the CSM + extended nickel mixing model may better match the near-linear rise. Analogously, extra power from collision of ejecta and unbound material could also improve the $^{56}$Ni-shell model.

\subsubsection{Color Comparison}\label{subsec:color_model}

\begin{figure}
    \centering
    \includegraphics[width=0.45\textwidth]{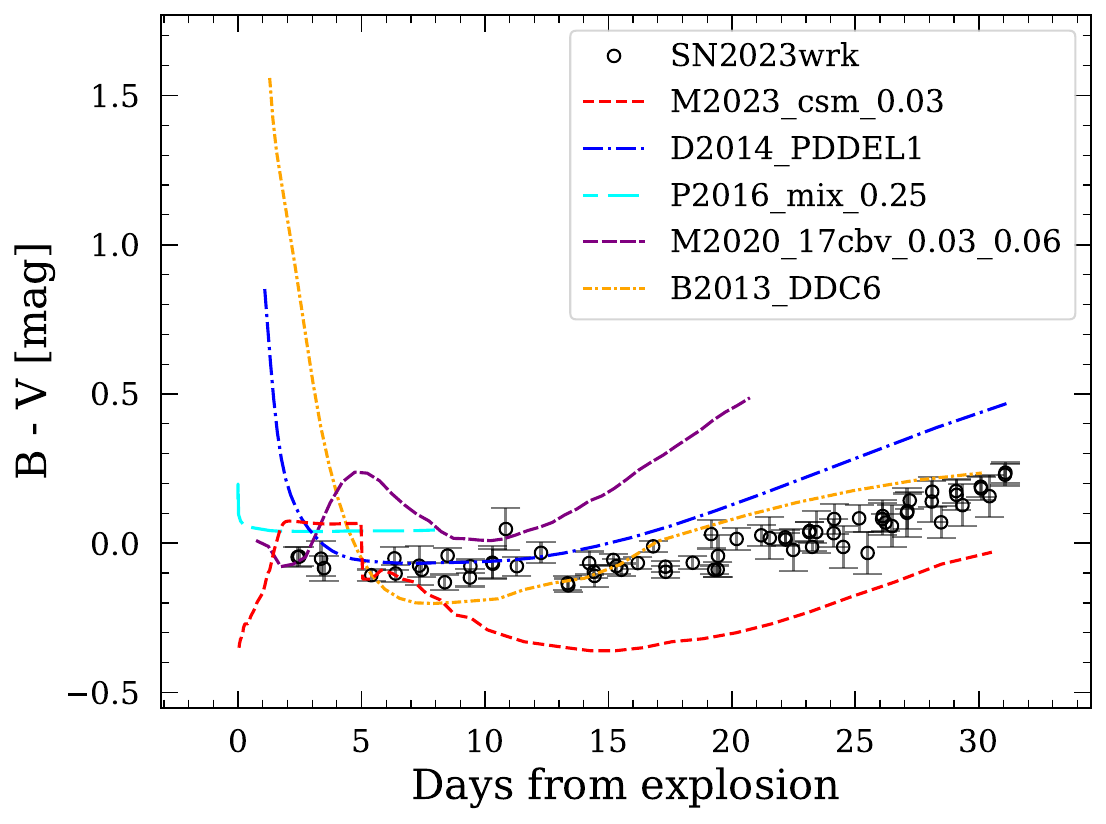}
    \caption{Comparison of early $B-V$ color between SN~2023wrk and the models presented in Figure\,\ref{fig:model_lc}. The $B-V$ color of SN~2023wrk has been corrected for reddening.
    }
    \label{fig:model_color}
\end{figure}

We compare the $B-V$ evolution between SN~2023wrk and some models in Figure\,\ref{fig:model_color}. As \citet{2018ApJ...852..100M} mentioned, the extended nickel mixing model is the only one that has a flat color evolution within days after explosion, though its $B-V$ color is $\sim 0.1$ mag redder than the observations. The $^{56}$Ni-shell model matches the earliest point well, but its color tends to get redder in the following days. Meanwhile, PDDEL1 predicts a blue and flat color evolution from $t\approx +4$ to +12 d after explosion. It seems that a combination of the $^{56}$Ni-shell and PDDEL1 models can match the observations better. At $t > 0$ d, the PDDEL1 model has a redder $B-V$ color than SN~2023wrk and thus a lower temperature. This can be seen clearly in the spectral comparison of SN~2023wrk and the PDDEL1 model near maximum light, where the spectrum of SN~2023wrk peaks at shorter wavelengths than that of PDDEL1 ($\sim 3500$~\AA\ vs. $\sim 4000$~\AA; see Figure\,\ref{fig:peak_SED}). This could not be attributed to different total masses of $^{56}$Ni, since the PDDEL1 model has a $^{56}$Ni mass of 0.76\,M$_{\odot}$ which is close to that of SN~2023wrk.

The $B-V$ color of the CSM interaction model is very blue at first and then evolves redward quickly, which is seen in neither SN~2023wrk nor iPTF16abc. Nevertheless, it is still possible that the observation missed such evolution, and the color is quickly dominated by the outer $^{56}$Ni. 
The one-dimensional standard delayed-detonation (DDT) model predicts a blueward color evolution within days after the explosion and thus is not consistent with SN~2023wrk, which could be improved by including the outward-mixing $^{56}$Ni. Note that the collision of unbound carbon and SN ejecta in the PDDEL models is similar to the carbon-rich CSM interaction to some extent \citep{2014MNRAS.441..532D}, so including the CSM interaction in DDT models could also improve the match with observations.   

\subsubsection{Spectral Comparison}\label{subsec:spec_model}

\begin{figure}
    \centering
    \includegraphics[width=0.45\textwidth]{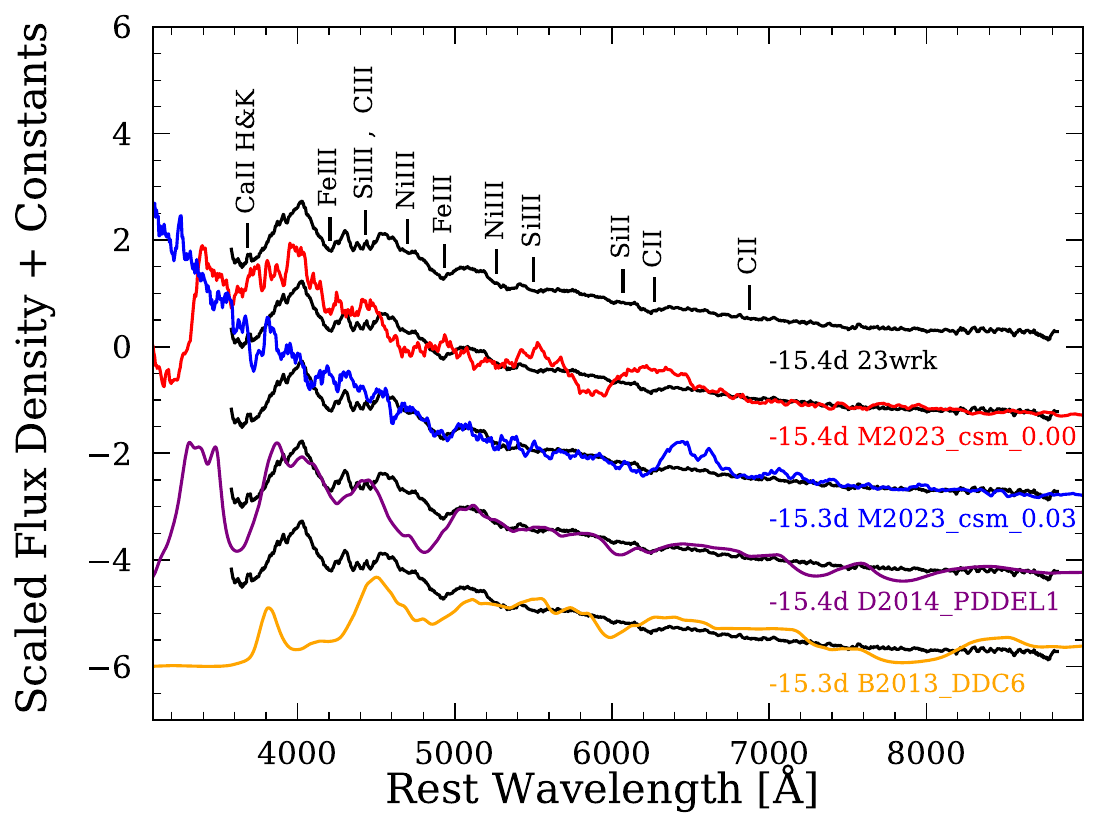}
    \caption{Spectral comparison between SN~2023wrk and synthetic spectra of some models at $t\approx -15$ d, including two super-$M_{\rm Ch}$ models without CSM interaction or with 0.03\, M$_{\odot}$ CSM interaction (M2023\_csm\_0.00 and M2023\_csm\_0.03; \citealp{2023MNRAS.521.1897M}), an $M_{\rm Ch}$ delayed-detonation model (DDC6; \citealp{2013MNRAS.429.2127B}), and an $M_{\rm Ch}$ pulsational delayed-detonation model (PDDEL1; \citealp{2014MNRAS.441..532D}). The spectrum of SN~2023wrk has been corrected for reddening and host redshift, and it has been smoothed with a window of 30~\AA. 
    }
    \label{fig:model_spec}
\end{figure}

We compare the $t\approx -15$ d spectra of SN~2023wrk and some models in Figure\,\ref{fig:model_spec}. Little unburned carbon is left in a standard DDT model, so the DDC6 model spectrum at $t\approx -15.3$ d shows no evident carbon absorption trough. The PDDEL model retains more unburned carbon owing to the pulsation, and thus prominent carbon features are seen in the spectrum of PDDEL1. In fact, the C\,{\sc ii} $\lambda6580$ lines of SN~2023wrk and PDDEL1 match well at $t\approx -15.4$ d. However, the PDDEL1 model has a clear Si\,{\sc ii} $\lambda6355$ line which is nearly absent in SN~2023wrk. \citet{2024MNRAS.529.3838A} find a silicon mass fraction of 0.013 at velocities $v>13,450$ km s$^{-1}$ for iPTF16abc, while the PDDEL1 model has a higher silicon mass fraction of $\sim 0.35$ in the same region. Thus, the weak silicon features of iPTF16abc at $t\approx -15.3$ d can be attributed to  silicon depletion in the outer layers. A similar case should hold for SN~2023wrk, as the spectra of these two SNe nearly overlap at this phase.  

Note that the PDDEL1 model has a weaker Si\,{\sc ii} line than the DDC6 model, though their silicon mass fractions in line-forming regions are similar. This could be due to a spike structure in the density profile (called ``cliff'' by \citealt{2014MNRAS.441..532D}), caused by collision of SN ejecta with the unbound carbon in the PDDEL models, while this is absent in the DDC models. When the photosphere recedes to the cliff, the large density gradient of the cliff reduces the spatial extent of the line formation and thus produces weak lines \citep{2005A&A...437..667D,2005A&A...439..671D}. This could explain the weak lines of SN~2023wrk.

The super-$M_{\rm Ch}$ model has a strong Si\,{\sc ii} $\lambda6355$ line at $t\approx -15.4$ d, but this feature is nearly absent because the photosphere is formed in the swept-up envelope when including a carbon-rich CSM. Thus, the collision of SN ejecta with nearby carbon could make the photosphere form in a higher-velocity layer and partially cause the nearly absent Si\,{\sc ii} line of SN~2023wrk at very early phases. Thus, the presence of carbon-rich CSM can account for prominent carbon features in the spectrum. Note that the collision of SN ejecta and carbon-rich CSM can also lead to a cliff in the density profile.

Unlike the continuous receding in the DDC model, the photosphere of the PDDEL model resides within the cliff for a longer time because it represents a large jump in optical depth \citep{2014MNRAS.441..532D}. Notice that the Si\,{\sc ii} velocity of SN 2023wrk is observed to have a plateau near $\sim 12,000$ km s$^{-1}$ during the period $t\approx -13.5$ to $-8.5$ d (see Figure\,\ref{fig:spec_velocities}), which may imply a cliff in density profile and thus supports a collision of SN ejecta and nearby material in SN~2023wrk. Moreover, it seems that such a cliff in SN~2023wrk is more prominent than that in PDDEL1, as the Si\,{\sc ii} velocity of SN~2023wrk has a flatter plateau and a faster decline after the plateau. 

In short, the PDDEL1 model gives a better match with SN~2023wrk, but some improvements are still needed. For instance, the luminosity is underestimated and the $B-V$ color is still too red at $t\lesssim +4$ d after explosion, which could be improved by including outward-mixing $^{56}$Ni. In fact, the Ni\,{\sc iii} features identified in the first spectrum of SN~2023wrk and the reappearance of carbon absorption around the time of $B$-band maximum light in an inner shell support the macroscopic mixing of fuel and ash from multidimensional effects as predicted by the three-dimensional DDT model \citep{2005ApJ...623..337G}, in which the unburned carbon and oxygen mix inward and the intermediate-mass elements (IMEs) and iron-group elements (IGEs) mix outward when averaging over angle. As PDDEL1 has a larger silicon mass fraction than SN~2023wrk in the outer layers, this could also be improved by including outward-mixing IGEs. Note that for the PDDEL models, a $^{56}$Ni shell in the outer layer is needed to reproduce the bump in the early light curve, though it could also lead to a very red peak color that is inconsistent with SN~2023wrk. An alternative choice is the DDT model with a carbon-rich CSM. In this case, the bump in the light curve could be produced by combining CSM interaction and the radioactive decay of outward-mixing $^{56}$Ni, and the blue and constant $B-V$ color evolution could arise from  extended $^{56}$Ni mixing. 

\subsection{Diversity among Carbon-rich SNe~Ia}\label{subsec:carbon-rich}

Besides iPTF16abc and SN\,2023wrk, there are increasing number of SNe~Ia with %comparable Si\,{\sc ii} $\lambda6355$ and C\,{\sc ii} $\lambda6580$ lines 
strong C\,{\sc ii} $\lambda6580$ absorption have been discovered, such as the normal objects SNe~2012cg, 2013dy, and 2017cbv, as well as 03fg-like or 02es-like SNe~Ia. Here we call these objects ``carbon-rich SNe~Ia''. Besides the prominent carbon features, these SNe~Ia show bluer colors and excess emission in optical and UV bands, and weaker Si\,{\sc ii} lines compared with the typical object SN~2011fe at early times. Except for 03fg-like or 02es-like objects, the light curves of carbon-rich SNe~Ia are homogeneous around maximum light (see Figure\,\ref{fig:LC_op_compare}), and we call them relatively normal carbon-rich SNe~Ia. Nevertheless, significant diversity is still found among these carbon-rich SNe~Ia. In this section, we briefly present this diversity and explore the possible origins.

Among our comparison sample, SN~2023wrk and iPTF16abc are more similar to 91T/99aa-like objects, since they have very weak Si\,{\sc ii} and prominent Fe\,{\sc iii} lines, whereas SNe~2012cg, 2013dy, and 2017cbv are closer to normal SNe~Ia, showing prominent Si\,{\sc ii} lines at early times. To measure the line-strength ratio of C\,{\sc ii} $\lambda6580$ to Si\,{\sc ii} $\lambda6355$, we determine the pseudo-equivalent width (pEW) ratio of these two lines by performing two-component Gaussian fits to them normalized by a pseudocontinuum which is defined as a straight line connecting the interactively chosen points on the red and blue sides of each feature. We compare the pEW ratios of SN~2023wrk and some other SNe~Ia with early-time observations in Figure\,\ref{fig:pEW_comp};
the individual pEWs measurements as well as the values of the ratio can be found in Appendix\,\ref{sec:pEW}. As can be seen, the pEW ratio of C\,{\sc ii} $\lambda6580$ to Si\,{\sc ii} $\lambda6355$ is higher for the 03fg-like objects and the 02es-like objects, but is smaller for the normal SNe~Ia and moderate for SN~2023wrk and iPTF16abc at comparable phases. If the unbound carbon is the main source of the strong carbon features, the diversity of the line-strength ratio may partially stem from different properties of the unbound carbon. For instance, more massive unbound carbon could lead to stronger and longer-lasting carbon features and probably weaker silicon lines owing to the formation of the photosphere above more of the silicon. Another difference is the early $B-V$ color evolution: it takes a longer time for SNe~2012cg, 2013dy, and 2017cbv to evolve to the blue, while SN~2023wrk and iPTF16abc have a constant $B-V$ color before maximum light. This may be due to different properties of the outward-mixing $^{56}$Ni, since a higher mass fraction of $^{56}$Ni in the outer layers can shorten the time to evolve blueward (see Figure 8 of \citealt{2016ApJ...826...96P}). In addition, it could explain the higher Fe\,{\sc iii}/Fe\,{\sc ii} ratios in the early spectra of SN~2023wrk and iPTF16abc.

\begin{figure}
    \centering
    \includegraphics[width=0.45\textwidth]{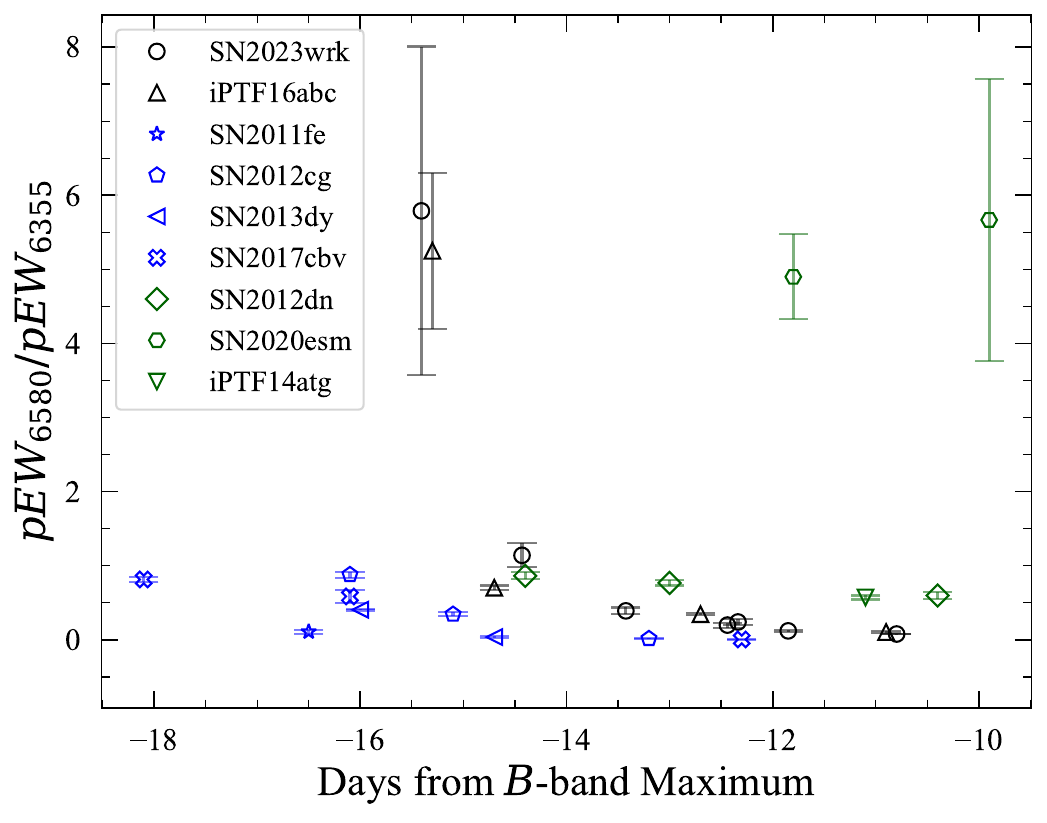}
    \caption{Comparison of evolution of the pEW ratio of C\,{\sc ii} $\lambda6580$ to Si\,{\sc ii} $\lambda6355$ in SN~2023wrk and SNe~2011fe, 2012cg, 2013dy, 2017cbv \citep{2017ApJ...845L..11H}, 2012dn \citep{2016MNRAS.457.3702P,2020MNRAS.492.4325S}, 2020esm \citep{2022ApJ...927...78D}, iPTF14atg \citep{2015Natur.521..328C}, and iPTF16abc.  
    91T/99aa-like SNe~Ia are shown in black, normal SNe~Ia are in blue, and 03fg-like or 02es-like objects are in dark green. The references for spectra of SNe~2011fe, 2012cg, 2013dy, and iPTF16abc can be found in the caption of Figure\,\ref{fig:spec_velocities}.
    }
    \label{fig:pEW_comp}
\end{figure}

Most of the 03fg-like or 02es-like SNe~Ia are more peculiar than other carbon-rich SNe~Ia and deviate from the Phillips relation (Figure 1 of \citealt{2017hsn..book..317T}). \citet{2024ApJ...966..139H} find that these objects have much bluer UV--optical colors than other SNe~Ia, which indicates little line blanketing by IGEs and thus probably little outward-mixing IGEs. These peculiar properties imply that different explosion mechanisms may exist in these objects and other carbon-rich SNe~Ia. For example, the violent merger (e.g., \citealp{2012ApJ...747L..10P}) of two C/O WDs is a popular model to explain the 03fg-like objects (e.g., SN~2022pul; \citealp{2024ApJ...960...88S}), where a detonation is triggered during the merger. 
Thus, in contrast to the three-dimensional DDT model, there are no outward-mixing IGEs through a deflagration in the violent merger.

There is also diversity among normal carbon-rich SNe~Ia. Within the first few days after explosion, SN~2013dy and SN 2017cbv are more luminous and slightly bluer than SN~2012cg, and their Si\,{\sc ii} $\lambda6355$ velocity is smaller than that of SN~2012cg. The pEW ratios of C\,{\sc ii} $\lambda6580$ to Si\,{\sc ii} $\lambda6355$ of these three SNe~Ia are similar at $t\approx -16$ d. But as seen in Figure\,\ref{fig:spec_compare}, the C\,{\sc ii} $\lambda6580$ and Si\,{\sc ii} $\lambda6355$ line profiles of SN~2012cg and SN~2013dy are also different. For instance, the absorption minimum of C\,{\sc ii} $\lambda6580$ has a smaller optical depth than that of Si\,{\sc ii} $\lambda6355$ in SN~2012cg, while the minima of these two lines have a similar optical depth in SN~2013dy. Also, the P~Cygni emission around 6500~\AA\ in SN~2012cg is stronger than that in SN~2013dy. Instead of the simple two-component Gaussian fits, more detailed analysis of the features is needed to accurately examine the difference. In addition, another normal object (SN~2018oh) is potentially a carbon-rich SN~Ia, since its C\,{\sc ii} $\lambda6580$ line at $t\approx -8.5$ d is still evident \citep{2019ApJ...870...12L}. Different from SNe~2012cg, 2013dy, and 2017cbv, the carbon features of SN~2018oh reappeared after $B$-band maximum. This difference could be attributed to a viewing-angle effect if the inner carbon comes from  asymmetric macroscopic mixing during the deflagration stage. But how the inner carbon survives in the subsequent detonation still needs to be explained.

It is important to study the progenitor system and explosion mechanism of the relatively normal carbon-rich SNe~Ia such as SN~2023wrk and SN~2013dy, since they could be used as standardizable candles and are usually more luminous than other normal SNe~Ia such as SN~2011fe. The diversity of the luminosity, color evolution, and absorption lines at very early times and the similarities around maximum light provide some clues for understanding their connections. However, the current sample of carbon-rich SNe~Ia is very limited, and many more carbon-rich SNe~Ia are needed to further explore their origins.

\section{CONCLUSION}\label{sec:conclusion}

In this paper, we present extensive photometric and spectroscopic observations of SN~2023wrk. It shows very close resemblances to iPTF16abc in many respects, including (1) evolution of early excess emission, (2) prominent but short-lived C\,{\sc ii} $\lambda6580$ absorption at early times and the reappearance from maximum light to three weeks thereafter, 
(3) a blue, constant $B-V$ color before maximum light, and (4) a location far away from the center of the host galaxy.   
After careful comparison with models, the early light-curve bump and the blue $B-V$ color seen in SN~2023wrk are found to be more consistent with the outer $^{56}$Ni-shell model proposed by \citet{2020A&A...642A.189M}. The subsequent evolution seems to be better matched by the PDDEL1 model \citep{2014MNRAS.441..532D}, though this model yields a redder SED around maximum light. The Ni\,{\sc iii} lines detected in very early spectra and the appearance of C\,{\sc ii} lines in post-peak spectra indicate macroscopic mixing of fuel and ash, as a result of multidimensional effects predicted by the three-dimensional DDT model. $^{56}$Ni mixing alone cannot explain the early linear rise in flux, which may be improved by including energy from the collision of SN ejecta with nearby material such as the 
unbound carbon in a CSM or PDDEL model. The carbon collision is supported by the strong carbon features in very early spectra and the $\sim 12,000$ km s$^{-1}$ plateau seen in the Si\,{\sc ii} $\lambda6355$ velocity evolution. A combination of a carbon-rich CSM and a $M_{\rm Ch}$ DDT model with strong $^{56}$Ni mixing may be an attractive model to interpret all the observations of SN~2023wrk; it needs to be explored in future work.

We find some diversity among carbon-rich SNe~Ia which have comparable C\,{\sc ii} $\lambda6580$ and Si\,{\sc ii} $\lambda6355$ lines in very early spectra, including the line-strength ratio of C\,{\sc ii} $\lambda6580$ to Si\,{\sc ii} $\lambda6355$, the iron absorption profile, the early-time luminosity, and the color evolution. Differences in properties of unbound carbon and outward-mixing $^{56}$Ni could partially account for this diversity. More carbon-rich SNe~Ia with very early observations are needed to further explore their connections and origins.


\section*{Acknowledgements}
This work is supported by the National Natural Science Foundation of China (NSFC grants 12288102, 12033003, and 12090044) and the Tencent Xplorer Prize. J.Z.L was supported by the Tianshan Talent Training Program through grant 2023TSYCCX0101. J.-J. Zhang and Y.-Z. Cai are supported by the International Centre of Supernovae, Yunnan Key Laboratory (No. 202302AN360001). J.-J. Zhang is supported by the National Key R\&D Program of China  (No. 2021YFA1600404), NSFC grant 12173082, the Yunnan Province Foundation (grant 202201AT070069), the Top-notch Young Talents Program of Yunnan Province, and the Light of West China Program provided by the Chinese Academy of Sciences. Y.-Z. Cai is supported by NSFC grant 12303054 and Yunnan Fundamental Research Projects grant  202401AU070063.
A.V.F.’s group at UC Berkeley received financial assistance from the Christopher R. Redlich Fund, as well as donations from Gary and Cynthia Bengier, Clark and Sharon Winslow, Alan Eustace, William Draper, Timothy and Melissa Draper, Briggs and Kathleen Wood, and Sanford Robertson (W.Z. is a Bengier-Winslow-Eustace Specialist in 
Astronomy, T.G.B. is a Draper-Wood-Robertson Specialist in Astronomy, Y.Y. was a Bengier-Winslow-Robertson Fellow in Astronomy), and numerous other donors. X.L. is supported by the Innovation Project of Beĳing Academy of Science and Technology (24CD013).
A.R. acknowledges financial support from the GRAWITA Large Program Grant (PI P. D'Avanzo) and the PRIN-INAF 2022 program ``Shedding light on the nature of gap transients: from the observations to the models.''
I.M. is grateful for support from Severo Ochoa grant CEX2021-001131-S funded by MCIN/AEI/ 10.13039/501100011033.

% Felipe's acknowledgement
The work of F.N. is supported by NOIRLab, which is managed by the Association of Universities for Research in Astronomy (AURA) under a cooperative agreement with the National Science Foundation.
% Wagner Corradi 
W. Corradi is supported by Laboratório Nacional de Astrofísica (LNA) through project numbers P-003, P-011, I-004, and I-013 from OPD. 
% Ted Leandro de Almeida
The work of L. de Almeida is supported by Laboratório Nacional de Astrofísica (LNA) through project numbers I-009, I-022, I-027, and P-008 from OPD. 
% Nelio Sasaki  
N. Sasaki is supported by Universidade do Estado do Amazonas -- UEA and LNA through project numbers P-003, P-008, I-004, and I-007.
%France
This work is supported by CNRS-MITI and Programme National des Hautes Energies (PNHE).

We thank the staffs at the various observatories where data were obtained.
A major upgrade of the Kast spectrograph on the Shane 3~m telescope at Lick Observatory, led by Brad Holden, was made possible through generous gifts from the Heising-Simons Foundation, William and Marina Kast, and the University of California Observatories. Research at Lick Observatory is partially supported by a generous gift from Google. 
Some of the data presented herein were obtained at the W. M. Keck Observatory, which is operated as a scientific partnership among the California Institute of Technology, the University of California, and NASA; the observatory was made possible by the generous financial support of the W. M. Keck Foundation.
The data presented here were obtained in part with ALFOSC, which is provided by the Instituto de Astrofisica de Andalucia (IAA-CSIC) under a joint agreement with the University of Copenhagen and NOT.
This work is partially based on observations collected with the Copernico 1.82~m telescope and the Schmidt 67/92 telescope (Asiago Mount Ekar, Italy) of the INAF -- Osservatorio Astronomico di Padova. We thank Stefan Taubenberger and his group of students from MPA for an epoch of ALFOSC imaging and a spectrum taken during a run for educational observations at the 1.82~m Copernico telescope. 
Part of the photometric data of this work were obtained by the Nanshan One-meter Wide-field Telescope which is supported by Tianshan Talent Training Program (grant 2023TSYCLJ0053) and the National Key R\&D program of China for Intergovernmental Scientific and Technological Innovation Cooperation Project under grant 2022YFE0126200.
We acknowledge the support of the staff of the Xinglong 80cm telescope (TNT). This work was partially supported by the Open Project Program of the Key Laboratory of Optical Astronomy, National Astronomical Observatories, Chinese Academy of Sciences. This research was partially based on observations made with the Thai Robotic Telescope under program ID TRTToO\_2024001 and  TRTC11B\_002, which is operated by the National Astronomical Research Institute of Thailand (Public Organization).
The ZTF forced-photometry service was funded under Heising-Simons Foundation grant 12540303 (PI M. Graham). This work makes use of the NASA/IPAC Extragalactic Database (NED), which is funded by NASA and operated by the California Institute of Technology.

%%TNG
This article is based in part on observations made at the Observatorios de Canarias del IAC with the TNG telescope operated on the island of La Palma by the Fundación Galileo Galilei INAF, Fundación Canaria (FGG), in the Observatorio del Roque de los Muchachos (ORM). 
%% FRAM
The operation of FRAM telescopes is supported by grants of the Ministry of Education of the Czech Republic LM2023032 and LM2023047, as well as EU/MEYS grants CZ.02.1.01/0.0/0.0/16\_013/0001403, CZ.02.1.01/0.0/0.0/18\_046/0016007,  CZ.02.1.01/0.0/0.0/ 16\_019/0000754, 
and CZ.02.01.01/00/22\_008/0004632.
%ADD FURTHER Acknowledgement from C. Adami (PLSS do not miss that). 
Based in part on observations made at Observatoire de Haute Provence (CNRS), France, with the MISTRAL instrument. This research has also made use of the MISTRAL database, operated at CeSAM (LAM), Marseille, France.
We acknowledge the excellent support from Observatoire de Haute-Provence, in particular Jerome Schmitt, Jean Claude Brunel, Francois Huppert, Jean Balcaen, Yoann Degot-Longhi, Stephane Favard, and Jean-Pierre Troncin.
We thank the OHP director for the allocations of two DDT observing slots.
% TRT
The work of K. Noysena and M. Tanasan is based on observations made with the Thai Robotic Telescope under program ID TRTC11A\_003, which is operated by the National Astronomical Research Institute of Thailand (Public Organization).
%IJCLAB
We acknowledge resources at IJCLAB for hosting science portals and data-reduction pipelines thanks to J. Peloton.

We acknowledge major contributions from the GRANDMA Collaboration and the Kilonova-Catcher citizen science program for the photometric and spectroscopic observations taken over 3 months and their subsequent data reduction. The authors express special gratitude to the PI of GRANDMA and the core team who decided to allocate substantial resources of the collaboration to the study of SN~2023wrk (S. Antier, D. Turpin, I. Tosta e Melo, M. Coughlin, P. Hello, P.-A. Duverne, S. Karpov, T. Pradier, J. Peloton, A. Klotz, C. Andrade).
%%Telescope team. 
We thank observers of the GRANDMA network program who actively participated in the follow-up campaign: AbAO observatory (N. Kochiashvili, R. Inasaridze and V. Aivazyan) for 1 observation; ShAO Observatory (N. Ismailov, Z. Vidadi, S. Aghayeva, S. Alishov, E. Hesenov) for 3 observations; HAO telescopes (A. Kaeouach) for 12 observations; FRAM telescopes (M. Mašek and S. Karpov) for 194 observations; KAO telescope (E. G. Elhosseiny, M. Abdelkareem, A. Takey, R. H. Mabrouk, A. Shokry, M. Aboueisha, Y. Hendy, A. E. Abdelaziz, R. Bendary, I. Zead, T. M. Kamel, G. M. Hamed, S. A. Ata, W. A. Badawy) for 67 observations;  TRT (K. Noysena, M. Tanasan) for 3 observations; and TBC, UBAI telescopes (Y. Tillayev, O. Burkhonov, Sh. Ehgamberdiev, Y. Rajabov) for 9 observations.
%
%% WC and Shifters
We acknowledge operations within GRANDMA with crucial participation of  22 follow-up advocates (C. Andrade, P. Gokuldass, M. Tanasan, I. Abdi, D. Akl, P. Hello, O. Pyshna, A. Simon, J-G. Ducoin, Y. Tillayev, M. Mašek, Z. Vidadi, Y. Rajabov, S. Antier, T. Leandro de Almeida, F. Navarete, N. Sasaki, W. Corradi, I. Tosta e Melo, M. Coughlin, T. Hussenot-Desenonges, S. Aghayeva) and their leaders (C. Andrade, P. Hello, S. Antier, I. Tosta e Melo), who participated in providing adequate observation strategies. 
%DAG
We thank the whole Data Analysis Group of GRANDMA consisting of I. Abdi, D. Akl, C. Andrade, J.-G. Ducoin, P.-A. Duverne, S. Karpov, F. Navarete, and D. Turpin.
P.-A. Duverne is grateful to Etienne Bertrand and Emmanuel Soubrouillard for providing the source's spectra and their help in analyzing these data.
%
%%KNC
D. Turpin, PI of the Kilonova-Catcher, acknowledges the observers of the Kilonova-Catcher program who actively participated in the follow-up campaign: E. Broens, F. Dubois, M. Freeberg, C. Galdies, R. Kneip, D. Marchais, E. Maris, R. Ménard, M. Odeh, G. Parent, A. Popowicz, D. ST-Gelais, and M. Serrau.
%ADD a sentence to Unistellar. 
We also thank Unistellar observers B. Guillet, B. Haremza, G. Di Tommaso, K. Borrot, M. Lorber, M. Shimizu, M. Mitchell, N. Meneghelli, N. Delaunoy, O. Clerget, P. Huth, P. Heafner, P. Kuossari, S. Saibi, S. Price, W. Ono, and Y. Arnaud.

%The authors acknowledge guidance for interaction with J. Liu and X. Wang's group to enhance the quality of the article represented by P. Hello and S. Antier.


\vspace{5mm}

\software{AUTOPHOT \citep{2022A&A...667A..62B}, HEASOFT(\url{https://www.swift.ac.uk/analysis/software.php}), Matplotlib \citep{Hunter2007}, NumPy \citep{2020Natur.585..357H}, Pandas \citep{mckinney-proc-scipy-2010}, SciPy \citep{2020NatMe..17..261V}, SNooPy2 \citep{2011AJ....141...19B}, STDPipe \citep{stdpipe} 
}



\appendix

\section{PHOTOMETRIC AND PSEUDOBOLOMETRIC LIGHT CURVES} \label{sec:all_LCs}

A summary of photometric observations triggered by us is presented in Table\,\ref{tab:obs-gm}, including telescope names and the corresponding filters and number of images. The full photometric data of SN~2023wrk are presented in Table\,\ref{table:ground_lc}. The estimated pseudobolometric light curve is presented in Table\,\ref{table:bolo_lc}.

\begin{table}[h!]
	\centering
  	\caption{Summary of photometric observations.}
	\begin{tabular}{|c | c | c|}
	\hline
	Facility & Number of images & Bands \\
	\hline\hline
    Atlas Sky Observatory (ASO) &  8 & L, g, r, i\\
    Abastumani-T70 & 1 & R \\
    FRAM-CTA-N & 194 & B, V, R\\
    HAO & 8 & R, V, g\\
    KAO & 67 & g, r, i, z\\
    Lisnyky-Schmidt-Cassegrain & 1 & R \\
    ShAO-2m & 3 & B, V, R \\
    TRT-SRO  & 3 & B, V\\
    UBAI-NT60 & 6 & B, V, R \\
    UBAI-ST60 & 3 & B, V, R \\
    \hline
    Kilonova-Catcher (17 amateur telescopes) & 337 & B, V, R, I, g, r, i\\
    \hline
    Copernico & 20 & B, V, u, g, r, i, z\\
    LJT & 48 & B, V, g, r, i, z\\
    NOWT & 102 & U, B, V, R, I\\
    TNT & 387 & B, V, g, r, i\\
    67/91-ST & 52 & B, V, u, g, r, i\\
    \hline
	\end{tabular}
	\label{tab:obs-gm}
    \end{table}

\begin{deluxetable}{lllll}
\caption{Observed photometry of SN~2023wrk}\label{table:ground_lc}
\tablehead{
\colhead{MJD} & \colhead{Magnitude} & \colhead{Error\tablenotemark{a}} & \colhead{Band} & \colhead{Source}
}
%MJD & Magnitude & Error$^{a}$ & Band & Source \\
\startdata
60251.524 & 20.621 & 0.323 & g & ZTF \\
60252.229 & 18.110 & 0.070 & L & GOTO \\
60253.467 & 16.905 & 0.051 & g & ZTF \\
60253.470 & 16.850 & 0.037 & g & ZTF \\
60253.497 & 16.930 & 0.011 & r & ZTF \\
60253.915 & 16.777 & 0.020 & B & LJT \\
60253.917 & 16.780 & 0.025 & V & LJT \\
60253.918 & 16.668 & 0.027 & g & LJT \\
60253.920 & 16.792 & 0.026 & r & LJT \\
60253.922 & 17.072 & 0.039 & i & LJT \\
60253.971 & 16.683 & 0.040 & I & NOWT \\
... & ... & ... & ... & ... \\
60366.826 & 17.835 & 0.045 & B & NOWT \\
60366.830 & 17.495 & 0.040 & V & NOWT \\
60366.834 & 17.549 & 0.027 & R & NOWT \\
60366.839 & 17.677 & 0.059 & I & NOWT \\
\enddata
\tablenotetext{a}{$1\sigma$.}
\tablecomments{This table is available in its entirety in machine-readable form.}
\end{deluxetable}

\begin{table*}
\centering
\caption{Estimated Bolometric Light Curve of SN~2023wrk}\label{table:bolo_lc}
\begin{tabular}{lllllllll}
\hline
\hline
Phase$^{a}$ & $L$ & Error$^{b}$ & Phase & $L$ & Error & Phase & $L$ & Error \\
Day   & (10$^{42}$ erg s$^{-1}$) & (10$^{42}$ erg s$^{-1}$) & Day   & (10$^{42}$ erg s$^{-1}$) & (10$^{42}$ erg s$^{-1}$) & Day   & (10$^{42}$ erg s$^{-1}$) & (10$^{42}$ erg s$^{-1}$) \\
\hline
-10.592 & 7.203 & 0.195 & 16.921 & 6.368 & 0.172 & 44.635 & 2.237 & 0.095 \\
-9.712 & 8.698 & 0.216 & 17.746 & 6.105 & 0.158 & 45.544 & 2.152 & 0.093 \\
-8.744 & 10.347 & 0.240 & 18.476 & 5.896 & 0.149 & 46.821 & 2.044 & 0.091 \\
-7.714 & 12.039 & 0.264 & 19.728 & 5.584 & 0.139 & 47.121 & 2.020 & 0.090 \\
-6.738 & 13.480 & 0.288 & 20.895 & 5.343 & 0.134 & 48.443 & 1.924 & 0.088 \\
-6.042 & 14.393 & 0.309 & 21.650 & 5.200 & 0.133 & 49.709 & 1.844 & 0.086 \\
-4.768 & 15.721 & 0.351 & 22.823 & 5.007 & 0.134 & 50.779 & 1.784 & 0.083 \\
-3.802 & 16.424 & 0.383 & 23.537 & 4.900 & 0.135 & 51.202 & 1.759 & 0.082 \\
-3.052 & 16.784 & 0.412 & 24.855 & 4.716 & 0.140 & 52.446 & 1.699 & 0.079 \\
-1.973 & 17.031 & 0.461 & 25.453 & 4.639 & 0.144 & 53.741 & 1.644 & 0.074 \\
-0.703 & 16.956 & 0.516 & 26.764 & 4.472 & 0.153 & 54.444 & 1.617 & 0.071 \\
-0.272 & 16.853 & 0.528 & 27.884 & 4.332 & 0.162 & 55.279 & 1.588 & 0.068 \\
1.106 & 16.318 & 0.521 & 28.695 & 4.231 & 0.167 & 56.531 & 1.547 & 0.065 \\
1.942 & 15.870 & 0.482 & 29.963 & 4.074 & 0.172 & 57.394 & 1.520 & 0.064 \\
3.252 & 15.038 & 0.408 & 30.658 & 3.988 & 0.171 & 58.771 & 1.478 & 0.064 \\
4.184 & 14.380 & 0.374 & 31.961 & 3.824 & 0.164 & 59.362 & 1.460 & 0.065 \\
5.202 & 13.624 & 0.349 & 32.831 & 3.714 & 0.156 & 60.212 & 1.433 & 0.066 \\
6.041 & 12.981 & 0.324 & 33.909 & 3.575 & 0.145 & 61.645 & 1.385 & 0.067 \\
7.053 & 12.205 & 0.284 & 34.418 & 3.509 & 0.140 & 62.453 & 1.355 & 0.067 \\
8.177 & 11.343 & 0.241 & 35.635 & 3.349 & 0.128 & 63.562 & 1.311 & 0.067 \\
9.128 & 10.631 & 0.220 & 36.827 & 3.189 & 0.119 & 64.493 & 1.281 & 0.070 \\
10.123 & 9.915 & 0.218 & 37.488 & 3.100 & 0.114 & 65.591 & 1.229 & 0.068 \\
11.094 & 9.258 & 0.226 & 38.849 & 2.918 & 0.108 & 66.506 & 1.182 & 0.065 \\
12.047 & 8.659 & 0.234 & 39.640 & 2.814 & 0.105 & 67.560 & 1.123 & 0.061 \\
13.139 & 8.031 & 0.235 & 40.620 & 2.689 & 0.103 & 68.098 & 1.091 & 0.058 \\
13.924 & 7.624 & 0.228 & 41.716 & 2.555 & 0.100 & 68.960 & 1.037 & 0.052 \\
15.377 & 6.956 & 0.204 & 42.744 & 2.436 & 0.098 & - & - & - \\
15.855 & 6.762 & 0.194 & 43.815 & 2.320 & 0.096 & - & - & - \\
\hline
\end{tabular}
\begin{flushleft}
  {$^a$ Relative to $B$-band maximum, MJD$_{B\rm max}=60269.45$.\\
  $^b$ Uncertainty in the distance not included. $1\sigma$.}  
\end{flushleft}
\end{table*}

\section{SPECTROSCOPIC OBSERVATIONS}\label{sec:all_spec}

The journal of spectroscopic observations of SN~2023wrk is presented in Table \ref{table:spec}.

\begin{table*}[h]
\centering
\caption{Overview of optical spectra of SN~2023wrk}\label{table:spec}
\begin{tabular}{llllll}
\hline
\hline
MJD & Date & Phase$^a$ & Range($\rm \AA$) & Exposure(s) & Instrument/Telescope \\
\hline
60253.9 & 20231105 & -15.4 & 3616-8925 & 2201 & YFOSC/LJT \\
60254.9 & 20231106 & -14.4 & 3772-8917 & 2700 & BFOSC/XLT \\
60255.9 & 20231107 & -13.4 & 3523-8925 & 2200 & YFOSC/LJT \\
60256.9 & 20231108 & -12.4 & 3518-8926 & 2200 & YFOSC/LJT \\
60257.0 & 20231108 & -12.3 & 3704-8182 & 300 & AFOSC/Copernico \\
60257.5 & 20231109 & -11.8 & 3636-10732 & 1800 & Kast/Shane \\
60258.5 & 20231110 & -10.8 & 3636-10734 & 1800 & Kast/Shane \\
60259.2 & 20231111 & -10.2 & 4111-8119 & 2700 & MISTRAL/OHP \\
60260.9 & 20231112 & -8.5 & 3769-8918 & 3000 & BFOSC/XLT \\
60262.9 & 20231114 & -6.5 & 3772-8919 & 3600 & BFOSC/XLT \\
60263.9 & 20231115 & -5.5 & 3772-8919 & 2400 & BFOSC/XLT \\
60265.2 & 20231117 & -4.2 & 4090-8062 & 4200 & MISTRAL/OHP \\
60265.9 & 20231117 & -3.5 & 3771-8918 & 2400 & BFOSC/XLT \\
60266.6 & 20231118 & -2.8 & 3152-10251 & 180 & LRIS/Keck~I \\
60268.3 & 20231120 & -1.2 & 4003-7867 & 300 & DOLoRES-LRS/TNG \\
60268.8 & 20231120 & -0.6 & 3777-8915 & 1800 & BFOSC/XLT \\
60270.9 & 20231122 & 1.4 & 3775-8914 & 2400 & BFOSC/XLT \\
60272.9 & 20231124 & 3.4 & 3772-8919 & 2700 & BFOSC/XLT \\
60275.9 & 20231127 & 6.3 & 3775-8917 & 3000 & BFOSC/XLT \\
60280.1 & 20231202 & 10.5 & 3926-7437 & 10800 & EBE \\
60280.8 & 20231202 & 11.3 & 3774-8915 & 3000 & BFOSC/XLT \\
60282.9 & 20231204 & 13.3 & 3774-8915 & 3000 & BFOSC/XLT \\
60285.0 & 20231206 & 15.4 & 3389-9283 & 900 & AFOSC/Copernico \\
60290.5 & 20231212 & 20.8 & 3640-8745 & 2100 & HIRES/Shane \\
60293.1 & 20231215 & 23.4 & 3816-7551 & 7200 & ESOU \\
60295.0 & 20231216 & 25.2 & 3501-9278 & 900 & AFOSC/Copernico \\
60295.0 & 20231216 & 25.3 & 4093-7434 & 7200 & EBE \\
60296.8 & 20231218 & 27.1 & 3777-8916 & 3600 & BFOSC/XLT \\
60313.9 & 20240104 & 44.0 & 3614-8925 & 1800 & YFOSC/LJT \\
60314.9 & 20240105 & 44.9 & 3775-8915 & 3000 & BFOSC/XLT \\
60316.0 & 20240106 & 46.0 & 3942-7444 & 10800 & EBE \\
60320.9 & 20240111 & 50.9 & 3996-7420 & 14400 & EBE \\
60321.4 & 20240112 & 51.4 & 3638-10756 & 1200 & Kast/Shane \\
60321.4 & 20240112 & 51.4 & 3637-8731 & 3600 & HIRES/Shane \\
60329.0 & 20240119 & 58.9 & 3988-7443 & 14400 & EBE \\
60330.0 & 20240121 & 60.0 & 3900-7200 & 7200 & ESOU \\
60332.1 & 20240123 & 62.0 & 4300-8121 & 2700 & MISTRAL/OHP \\
60380.3 & 20240311 & 109.7 & 3622-10754 & 3000 & Kast/Shane \\
\hline
\end{tabular}
\begin{flushleft}
\centering
{$^a$ Relative to the $B$-band maximum light, MJD$_{B\rm max}=60269.45$.}
\end{flushleft}
\end{table*}

\section{Parameters of comparison SNe~Ia in this Work} \label{sec:other_para}

The adopted distance modulus, total reddening, time of first light, and the references of the comparison SNe~Ia in this work are presented in Table\,\ref{table:other_para}. 

\begin{deluxetable}{llclcl}
\tablecaption{Parameters and references of the comparison SNe~Ia  }\label{table:other_para}
\tablehead{
\colhead{Name} & \colhead{$\mu$\tablenotemark{a}} & \colhead{$E(B-V)$} & \colhead{$t_0$\tablenotemark{b}} & \colhead{Early Excess?} & \colhead{References} \\
\nocolhead{Name} & \colhead{mag} & \colhead{mag} & \nocolhead{$t_0$} & \nocolhead{Early Excess?} & \nocolhead{References} 
}
\startdata
SN 2011fe  & 29.04$\pm$0.08 & 0.04 & 55796.84 & No & 1,2,3 \\
SN 2012cg  & 30.91$\pm$0.30 & 0.2 & 56062.7  & Yes & 2,4,5 \\
SN 2013dy  & 31.50$\pm$0.08 & 0.3 & 56482.99 & Yes & 2,6,7, \\
SN 2017cbv & 30.58$\pm$0.05 & 0.145 & 57821.9  & Yes & 2,8,9 \\
SN 2018oh  & 33.61$\pm$0.05 & 0.038 & 58144.3 & Yes & 2,10,11,12\\
iPTF16abc  & 35.08$\pm$0.15 & 0.078 & 57481.21 & Yes & 2,13,14,15 \\
\enddata
\tablenotetext{a}{Distance modulus.}
\tablenotetext{b}{Time of first light (MJD).}
\tablecomments{References: (1) \citealt{2001ApJ...549..721M}; (2) \citealt{2011ApJ...737..103S}; (3) \citealt{2016ApJ...820...67Z}; (4) \citealt{2008ApJ...683...78C}; (5) \citealt{2012ApJ...756L...7S}; (6) \citealt{2016ApJ...826...56R}; (7) \citealt{2013ApJ...778L..15Z}; (8) \citealt{2020ApJ...904...14W}; (9) \citealt{2017ApJ...845L..11H}; (10) \citealt{2019ApJ...870...12L}; (11) \citealt{2019ApJ...870...13S}; (12) \citealt{2019ApJ...870L...1D}; (13) \citealt{2000ApJ...529..786M}; (14) \citealt{2017AA...606A.111F}; (15) \citealt{2018ApJ...852..100M}.}
\end{deluxetable}

\section{Results of pEW measurements} \label{sec:pEW}

The pEWs of C\,{\sc ii} $\lambda6580$ and Si\,{\sc ii} $\lambda6355$ measured by the two-component Gaussian fits as well as their ratios are presented in Table\,\ref{table:pEW}. 

\begin{deluxetable}{lcccc}
\tablecaption{Individual
pEWs measurements as well as the values of the ratio}\label{table:pEW}
\tablehead{
\colhead{Name} & \colhead{Phase} & \colhead{pEW(CII 6580 \AA)} & \colhead{pEW(SiII 6355 \AA)} & \colhead{pEW(CII 6580 \AA)/pEW(SiII 6355 \AA)} \\
\nocolhead{Name} & \colhead{Day} & \colhead{\AA} & \colhead{\AA} & \nocolhead{pEW(CII 6580 \AA)/pEW(SiII 6355 \AA)} 
}
\startdata
SN2023wrk & -15.4 & 22.17$\pm$1.37 & 3.83$\pm$1.45 & 5.79$\pm$2.22 \\
SN2023wrk & -14.4 & 17.34$\pm$1.64 & 15.20$\pm$1.60 & 1.14$\pm$0.16 \\
SN2023wrk & -13.4 & 10.68$\pm$1.16 & 27.29$\pm$1.26 & 0.39$\pm$0.05 \\
SN2023wrk & -12.4 & 6.51$\pm$0.93 & 32.65$\pm$2.92 & 0.20$\pm$0.03 \\
SN2023wrk & -12.3 & 7.33$\pm$1.00 & 30.39$\pm$1.55 & 0.24$\pm$0.04 \\
SN2023wrk & -11.8 & 3.73$\pm$0.29 & 30.79$\pm$1.08 & 0.12$\pm$0.01 \\
SN2023wrk & -10.8 & 2.30$\pm$0.16 & 29.10$\pm$0.56 & 0.08$\pm$0.01 \\
iPTF16abc & -15.3 & 18.85$\pm$0.56 & 3.59$\pm$0.71 & 5.25$\pm$1.05 \\
iPTF16abc & -14.7 & 13.14$\pm$0.37 & 18.67$\pm$0.60 & 0.70$\pm$0.03 \\
iPTF16abc & -12.7 & 8.99$\pm$0.35 & 25.92$\pm$0.82 & 0.35$\pm$0.02 \\
iPTF16abc & -10.9 & 3.30$\pm$0.41 & 31.32$\pm$0.82 & 0.11$\pm$0.01 \\
SN2011fe & -16.5 & 11.97$\pm$2.56 & 109.47$\pm$6.28 & 0.11$\pm$0.02 \\
SN2012cg & -16.1 & 59.97$\pm$1.97 & 68.20$\pm$2.14 & 0.88$\pm$0.04 \\
SN2012cg & -15.1 & 35.73$\pm$2.56 & 103.82$\pm$3.01 & 0.34$\pm$0.03 \\
SN2012cg & -13.2 & 1.68$\pm$0.49 & 102.99$\pm$1.28 & 0.02$\pm$0.00 \\
SN2013dy & -16.0 & 28.53$\pm$0.52 & 70.20$\pm$1.16 & 0.41$\pm$0.01 \\
SN2013dy & -14.7 & 3.13$\pm$0.72 & 82.92$\pm$1.80 & 0.04$\pm$0.01 \\
SN2017cbv & -18.1 & 43.47$\pm$1.34 & 53.33$\pm$1.59 & 0.82$\pm$0.04 \\
SN2017cbv & -16.1 & 34.97$\pm$4.58 & 59.69$\pm$4.26 & 0.59$\pm$0.09 \\
SN2017cbv & -12.3 & 0.44$\pm$0.31 & 78.39$\pm$1.34 & 0.01$\pm$0.00 \\
SN2012dn & -14.4 & 3.52$\pm$0.18 & 4.07$\pm$0.08 & 0.86$\pm$0.05 \\
SN2012dn & -13.0 & 27.13$\pm$1.07 & 35.31$\pm$0.80 & 0.77$\pm$0.03 \\
SN2012dn & -10.4 & 2.19$\pm$0.18 & 3.66$\pm$0.12 & 0.60$\pm$0.05 \\
SN2020esm & -11.8 & 72.46$\pm$2.22 & 14.79$\pm$1.67 & 4.90$\pm$0.57 \\
SN2020esm & -9.9 & 50.25$\pm$2.80 & 8.87$\pm$2.94 & 5.67$\pm$1.91 \\
iPTF14atg & -11.1 & 20.30$\pm$0.81 & 35.50$\pm$0.91 & 0.57$\pm$0.03 \\
\enddata
\end{deluxetable}


\bibliography{sn-bibliography}{}
\bibliographystyle{aasjournal}



\end{document}